\renewcommand\L{\Lambda}

\newcommand{\diracslash}[1]{#1\llap{/\kern2pt}}

\newcommand{\be}{\begin{equation}}
\newcommand{\ee}{\end{equation}}
\newcommand{\bea}{\begin{eqnarray}}
\newcommand{\eea}{\end{eqnarray}}
\newcommand{\ba}[1]{\begin{array}{#1}}
\newcommand{\ea}{\end{array}}

\newcommand{\bt}{\begin{tabular}}
\newcommand{\et}{\end{tabular}}

\newcommand{\beas}{\begin{eqnarray*}}
\newcommand{\eeas}{\end{eqnarray*}}

\documentclass[preprint,prd,aps,floats,nofootinbib,floatfix]{revtex4}

\DeclareSymbolFont{rsfs}{U}{rsfs}{m}{n}
\DeclareSymbolFontAlphabet{\mathrsfs}{rsfs}
\usepackage{graphicx}

\usepackage{multirow}

\usepackage{aliascnt}

\usepackage{pgf}
\usepackage{csquotes}
\usepackage{pdfpages}
\usepackage{amsmath}
\usepackage[amsmath,thmmarks]{ntheorem}
\usepackage{epstopdf}

\usepackage{textcomp}

\usepackage{graphicx}

\usepackage{cleveref}

\begin{document}

\title{In-medium properties of pseudoscalar $D_s$ and $B_s$ mesons}

\author{Rahul Chhabra}
\email{rahulchhabra@ymail.com,}
\affiliation{Department of Physics, Dr. B R Ambedkar National Institute of Technology Jalandhar,
 Jalandhar -- 144011,Punjab, India}

\author{Arvind Kumar}
\email{iitd.arvind@gmail.com, kumara@nitj.ac.in}
\affiliation{Department of Physics, Dr. B R Ambedkar National Institute of Technology Jalandhar,
 Jalandhar -- 144011,Punjab, India}

\def\be{\begin{equation}}
\def\ee{\end{equation}}
\def\bearr{\begin{eqnarray}}
\def\eearr{\end{eqnarray}}
\def\zbf#1{{\bf {#1}}}
\def\bfm#1{\mbox{\boldmath $#1$}}
\def\hf{\frac{1}{2}}
\def\kp{\zbf k+\frac{\zbf q}{2}}
\def\km{-\zbf k+\frac{\zbf q}{2}}
\def\hwo{\hat\omega_1}
\def\hwt{\hat\omega_2}

\begin{abstract}

We calculate the shift in masses and decay constants of $D_s(1968)$ and $B_s(5370)$ mesons in hot and dense asymmetric strange hadronic matter using  QCD sum rules and chiral SU(3) model. In-medium strange quark condensates $\left\langle \bar{s}s\right\rangle_{\rho_B}$, and gluon condensates $\left\langle  \frac{\alpha_{s}}{\pi} {G^a}_{\mu\nu} {G^a}^{\mu\nu}
\right\rangle_{\rho_B}$ to be used in the QCD sum rules for pseudoscalar  $D_s$ and $B_s$ mesons are calculated using chiral SU(3) model. As an application of our present work, we calculate the in-medium decay widths of the excited (c$\bar{s}$) states  $D_s^*(2715)$ and $D_s^*(2860)$  decaying to ($D_s(1968)$,$\eta$) mesons. The medium effects in their decay widths are incorporated through the mass modification of the $D_s(1968)$ and $\eta$ mesons. Results of the present investigation may be helpful to understand the possible outcomes of the future experiments like CBM and PANDA under the FAIR facility. 


\end{abstract}

\maketitle
\section{Introduction}

Exploring the QCD phase diagram at different temperatures and baryon chemical potentials is an interesting and challenging problem. 
The heavy-ion collision experiments e.g., RHIC and LHC are intended to probe the phases of QCD at a region of high temperature and low baryonic density. However, the future facilities like CBM and PANDA of the FAIR project (at GSI Germany) are expected to explore the QCD phases at a region of high baryonic density and moderate temperature.   The possibility of the production of open charm 
in these heavy ion collision experiments and also in the J-PARC facility, encourage us to investigate heavy mesons in nuclear matter \cite{babar,J}. In particular, the study of the open charm  and bottom mesons in nuclear as well as in strange hadronic matter, can highlight the important points about the yield of $J/\psi$ and $\Upsilon$ states produced in the heavy ion collision experiments. It was first predicted by Matsui \cite{mats}, that the decrease in the yield of $J/ \psi$ state in the medium, should be considered as a signature for the production of quark gluon plasma in HIC experiments. 
In Ref. \cite{bochk} authors observed drastic change in the effective spectrum of vector channel $J^P=1^{++}$ at temperature interval $\simeq$ 150-200 MeV, in hadronic matter. Also, using sum rule approach with the conservative stability
criteria, authors did not observe drastic effect of dynamical
fermions on phase transitions in QCD \cite{dosch}.  Further, in Ref. \cite{pwave}(\cite{swave}) auhors observed survival of $\eta_c$ and $\chi_c$($\Upsilon$ and $\eta_b$) states beyond the critical temperature $T_c$. On the other hand, $\chi_b$ and $h_b$ states were observed to melt in QGP phase in \cite{aarts1}.   

Moreover,  the experimental evidences of $J/\psi$ suppression were observed by the collaborations NA38 \cite{NA38}, NA50 \cite{NA50pb} and  NA60 \cite{NA60}. 
The results in the favour of $J/\psi$ suppression  were also observed
in the RHIC experiment \cite{rhic}, whereas the decrease in the yield of hidden bottom mesons ($\Upsilon(1S), \Upsilon(2S)$, etc) was observed in the  Pb-Pb collisions of LHC experiment \cite{upsilon}. However,  in \cite{gore,cassing1, cassing2}, authors claimed that, the in-medium modification of $D$ meson can  also alter the yield of $J/\psi$ in HIC experiments. One expects that, if the masses of $D$ ($B$) mesons decrease in the medium then the higher charmonium (bottomonium) states may decay to $D$ $(B)$ mesons instead of $J/\psi$ ($\Upsilon$) and hence this may decrease the yield of $J/\psi$ ($\Upsilon$) in the hadronic medium as well.
 Therefore, to avoid erroneously consideration of hadronic phase as QGP phase, the in-medium study of $D$ and $B$  mesons become  important. 
 In addition to $J/ \psi$ ($\Upsilon$) suppression, study of $D$ ($B$) mesons in hadronic medium may also reveal about the existence of the bound states of $D$ ($B$) mesons with nucleons \cite{tsu}, as well as with hyperons \cite{hoff}. Furthermore, to understand the results of enhanced production of $D$ and $\bar{D}$ mesons  in the antiproton-nucleon collision  the study of in-medium  properties of $D$ mesons is important \cite{alice,sibi}. The detailed recent review on the study of experimental and theoretical progress of open charmed and bottom mesons can be found in the paper \cite{review}.

  The study of  ground states $D_s (c\bar{s})$ and  $B_s (\bar{b}s$) mesons in hot and dense asymmetric hadronic matter may help to understand the in-medium interactions of light and heavy quarks, and this further may also explain the diffusion and hadronization by the efficacy of strangeness enhancement in the ultrarelativistic heavy ion collision experiments \cite{he,rapp}. Also, the comparison of elliptic flow of the in-medium strange $D_s$ mesons with the non-strange $D$ mesons, may help to analyse quantitatively the hadronic transport coefficients, and this  further helps in quantitative understanding of  viscosity to entropy ratio in hadronic matter \cite{he}.
   Further, the knowledge of the in-medium properties of $D_s$ and $B_s$ mesons help to understand the in-medium leptonic decay constants and this may further, lead to the understanding of heavy flavour electroweak transition and CP violation. In the chiral limit, we have $\frac{f_{D_s}}{f_{D}}=1$ and $\frac{f_{B_s}}{f_{B}}=1$, and when chiral symmetry breaks, there ratio deviate unity \cite{domi,nar1994,nar2013,NaD,baza}. For example, using lattice QCD, the ratios were predicted as $\frac{f_{D_s}}{f_{D}}=1.188$ and $\frac{f_{B_s}}{f_{B}}=1.229$ \cite{baza}. Similarly, in \cite{NaD}, the value of $\frac{f_{D_s}}{f_{D}}$ was observed as 1.187. Here we point out that, as the in-medium decay constants of $D_s/B_s$ and $D/B$ mesons may behave differently in the hadronic medium and therefore it will be of interest to understand the extent of flavour symmetry breaking effect in future heavy ion collision experiments. Recently, an enhanced ratio $\frac{D_s^+}{D^0}$ was measured in Au-Au collision at $\sqrt{s_{NN}}$=200 GeV in STAR experiment \cite{star}.
 
Theoretically, many methodologies have been developed to study the in-medium $D$ and $B$ mesons, giving different results. For example, in Ref. \cite{tsu}, authors observed a drop of the mass of $D$ meson using Quark Meson Coupling Model (QMC), in which quarks and gluons are degrees of freedom, and interactions between $D$ mesons and nucleons are considered through the exchange of scalar and vector mesons. Another approach is self-consistent coupled channel, which considers the hadrons as degrees of freedom \cite{tolo}. This approach was further modified from SU(3) flavour \cite{tolo}, to SU(4) and breaking of SU(4) symmetry via exchange of vector mesons \cite{hoff, lutz1}.
Using this approach authors calculated the positive shift in the mass of $D$ mesons \cite{tolo1}, whereas the negative shift in mass of $D_s$ mesons in strange hadronic matter was observed \cite{hoff, CE, CE1}.

 On the other hand,  in the QCD sum rules, operator product expansion (OPE) is applied on the current-current correlation function and Borel transformation is used to  equate the mass dependent terms and the various condensates up to dimension four \cite{haya2000}, further modified to dimension five \cite{hilger2009}. Using the linear density approximate QCD sum rules, author separate the even and odd terms of the correlation function to evaluate the mass splitting between $D$ and $\bar{D}$ mesons, as well as between $D_s$ and $\bar{D_s}$ mesons\cite{hilger2009}. 
In \cite{Rahul} using QCD sum rules and chiral SU(3) model, we evaluated the in-medium masses and decay constants of vector and axial vector $D_1$ and $B_1$ mesons, and observed repulsive interaction for axial vector whereas, attractive interaction for vector $D^*$ and $B^*$ mesons in hadronic medium \cite{Rahul}. Recently, in ref. \cite{suzu} authors used the Gaussian transformed QCD sum rules with Maximum Entropy Method to calculate  the mass splitting between $D$ and $\bar{D}$ mesons, followed by the Bayesian approach \cite{suzur}. Another model based on the chiral SU(3) symmetry had been widely used in the past to calculate in-medium mass of hadrons \cite{mishraB} and properties of the neutron stars \cite{mishraA}. In \cite{Dmishra} authors observed decrease in the mass of light vector mesons including the baryon Dirac sea, through the summation of baryon tadpole diagram in relativistic hartree approximation. Using chiral SU(3) model authors studied the masses and optical potential of kaons and antikaons in the nuclear matter \cite{mishraE, mishraS}, and also in hyperonic matter \cite{mishra_asym}. To observe the behavior of open charmed and bottom mesons in nuclear matter the  SU(3) model was generalised to SU(4) and SU(5) sector, and negative shift in the masses of $D$ and $B$ meson was observed \cite{su4, chiral4, chiral41}.   In \cite{pathak1, pathak2}, authors used the same approach of generalizing the SU(3) model to SU(4), in order to investigate the interaction of $D_s$ and $B_s$ mesons, in asymmetric strange hadronic matter at finite temperature.  Using this approach authors calculate the attractive interaction of $D_s$/$B_s$ mesons in hot and dense asymmetric hyperonic (along with the nucleons) medium.

    In the present investigation, we apply the chiral SU(3) model followed by the QCD sum rules approach, to calculate the shift in masses and decay constants of pseudoscalar strange charmed ($D_s$) and bottom ($B_s$) mesons in strange hadronic matter.  We first calculate the strange quark condensates $\left\langle \bar{s}s\right\rangle$,  and gluon condensates $\left\langle  \frac{\alpha_{s}}{\pi} {G^a}_{\mu\nu} {G^a}^{\mu\nu}
\right\rangle$, in hot and dense asymmetric hadronic matter through chiral SU(3) model,  and then use these condensates as input in the Borel transformed QCD sum rules to find the  medium modified masses and decay constants for $D_s$ and $B_s$ mesons.  
Furthermore, as an application of the mass modification of $D_s(1968)$ meson, we investigate the in-medium two mode partial decay width of $D_s^*(2715)$ and $D_s^*(2860)$ states decaying to ($D_s(1968),\eta$) mesons, using $^3 P_0$ model.  Additionally, we introduce the in-medium mass of $\eta$ meson, calculated using the heavy baryon chiral perturbation model along with the relativistic mean field theory \cite{zhong1}.

The $^3P_0$ model had been applied in the past  to compute the values of two body strong decay widths of various mesons \cite{micu,yao,barn1,barn2,close,sego,ferre1,ferre2,ferre3,close1,zhong,chen,Li1}. Two states  $D_s^*(2715)$ and $D_s^*(2860)$ are of particularly importance as, newly predicted by the Belle and Babar collaboration, with full decay widths nearly $\Gamma$ = 115 and 48 MeV, respectively. However the clear cut assignment of the quantum numbers is still  not uniquely confirmed. Theoretically, authors had observed various decay modes (in vacuum) of $D_s^*(2715)$ and $D_s^*(2860)$ mesons to judge their exact quantum numbers. For example, in Ref. \cite{liu,liu1,close,close1,godfrey}, authors used $^3P_0$ model to investigate different partial decay widths of above mentioned  mesons and authors suggested, the possible quantum numbers of $D_s^*(2715)$ and $D_s^*(2860)$ mesons, as $1^-(1^3D_1)$ and $3^-(1^3D_3)$, with the slight possibility of the states to be assigned as, $1^-(2^3S_1)$ and $1^-(1^3D_1)$ respectively. As mentioned in \cite{liu}, the observed decay modes of $D_s^*(2715)$ and $D_s^*(2860)$ as $D_s(1968) + \eta$ will unveil the finite possibility (though small) to assign their states as $1^-(2^3S_1)$ and $1^-(1^3D_1)$ respectively. We consider  this statement of particular interest, and try to impose the medium effects on the partial decay widths of $D_s^*(2715)$ and $D_s^*(2860)$ states decaying to ($D_s(1968), \eta$), and 
will concentrate on the possible shift in the partial decay widths.

 The outline of this paper is as follows: In \cref{chiral}, we describe the chiral SU(3) model to calculate in-medium strange quark and gluon condensates. In \cref{QCD}, we discuss the QCD sum rules to investigate the in-medium masses and decay constants of $D_s$ and $B_s$ mesons. In \cref{3P0}, we narrate the $^3P_0$ model, which will be used to calculate the in-medium partial decay widths of $D_s^*(2715)$ and $D_s^*(2860)$ mesons decaying to ($D_s(1968),\eta$).  In \cref{results}, we discuss the various results of the present work and finally in \cref{summary}, we shall summarize the present work. 
  \section{Chiral SU(3) model}
  \label{chiral}
The chiral SU(3) model is an effective theory based on the chiral property of the quarks $(m_u=m_d=m_s=0)$ i.e, invariance under the chiral transformation. In the chiral SU(3) model we start with the effective Lagrangian density which contains  the kinetic energy term, baryon meson interaction term which produce baryon mass, self-interaction of vector mesons which generates the dynamical mass of vector mesons, scalar mesons interactions which induce the spontaneous  breaking of chiral symmetry, and the explicit breaking term of chiral symmetry. It is based on the non-realization of chiral symmetry, broken scale invariance with the spontaneous breaking of
chiral symmetry properties 
 \cite{papa_nuclei}. In the present investigation, we use the mean field approximation to solve the effective Lagrangian density and under this scheme we replace the quantum field operator by their classical expectation values. Further, from Lagrangian density and using Euler Lagrange equation of motion, $\frac{\partial \mathcal{L}}{\partial \phi}-\partial_\mu (\frac{\partial \mathcal{L}}{\partial(\partial_\mu \phi)})=0$ where $\phi$ is scalar field, we obtain coupled equations of motion for the scalar fields $\sigma$, $\zeta$, $\delta$  and scalar dilaton field $\chi$, given as \cite{ss,arvind2}
\begin{align}
& k_{0}\chi^{2}\sigma-4k_{1}\left( \sigma^{2}+\zeta^{2}
+\delta^{2}\right)\sigma-2k_{2}\left( \sigma^{3}+3\sigma\delta^{2}\right)
-2k_{3}\chi\sigma\zeta \nonumber\\
-&\frac{d}{3} \chi^{4} \bigg (\frac{2\sigma}{\sigma^{2}-\delta^{2}}\bigg )
+\left( \frac{\chi}{\chi_{0}}\right) ^{2}m_{\pi}^{2}f_{\pi}
-\sum g_{\sigma i}\rho_{i}^{s} = 0,
\label{sigma}
\end{align}
\begin{align}
& k_{0}\chi^{2}\zeta-4k_{1}\left( \sigma^{2}+\zeta^{2}+\delta^{2}\right)
\zeta-4k_{2}\zeta^{3}-k_{3}\chi\left( \sigma^{2}-\delta^{2}\right)\nonumber\\
-&\frac{d}{3}\frac{\chi^{4}}{\zeta}+\left(\frac{\chi}{\chi_{0}} \right) 
^{2}\left[ \sqrt{2}m_{K}^{2}f_{K}-\frac{1}{\sqrt{2}} m_{\pi}^{2}f_{\pi}\right]
 -\sum g_{\zeta i}\rho_{i}^{s} = 0,
\label{zeta}
\end{align}
\begin{align}
 & k_{0}\chi^{2}\delta-4k_{1}\left( \sigma^{2}+\zeta^{2}+\delta^{2}\right)
\delta-2k_{2}\left( \delta^{3}+3\sigma^{2}\delta\right) +k_{3}\chi\delta 
\zeta \nonumber\\
 + &  \frac{2}{3} d \chi^4 \left( \frac{\delta}{\sigma^{2}-\delta^{2}}\right)
-\sum g_{\delta i}\rho_{i}^{s} = 0,
\label{delta}
\end{align}
 
\begin{align}
 & k_{0}\chi \left( \sigma^{2}+\zeta^{2}+\delta^{2}\right)-k_{3}
\left( \sigma^{2}-\delta^{2}\right)\zeta + \chi^{3}\left[1
+{\rm {ln}}\left( \frac{\chi^{4}}{\chi_{0}^{4}}\right)  \right]
+(4k_{4}-d)\chi^{3}
\nonumber\\
 - & \frac{4}{3} d \chi^{3} {\rm {ln}} \Bigg ( \bigg (\frac{\left( \sigma^{2}
-\delta^{2}\right) \zeta}{\sigma_{0}^{2}\zeta_{0}} \bigg ) 
\bigg (\frac{\chi}{\chi_0}\bigg)^3 \Bigg )\nonumber\\ 
 + &\frac{2\chi}{\chi_{0}^{2}}\left[ m_{\pi}^{2} f_{\pi}\sigma +\left(\sqrt{2}m_{K}^{2}f_{K}-\frac{1}{\sqrt{2}}
m_{\pi}^{2}f_{\pi} \right) \zeta\right]  = 0,
\label{chi}
\end{align}
respectively.
In the above equations, $m_K$ and $f_K$($m_{\pi}$ and $f_{\pi}$) denote the mass and decay constant of $K$($\pi$) meson, respectively and  the other parameters $k_0, k_1, k_2$, $k_3$ and $k_4$ are fitted  so as to reproduce the vacuum masses of $\eta$ and  $\eta'$  mesons \cite{papa_nuclei}. Further, ${\rho_i}^s$ represents the scalar density for $i^{th}$ baryon ($i=p, n,$ $\L$, $\Sigma^{\pm,0}$, $\Xi^{-,0}$) and is defined as 
\begin{align}
\rho_{i}^{s} = \gamma_{i}\int\frac{d^{3}k}{(2\pi)^{3}} 
\frac{m_{i}^{*}}{E_{i}^{*}(k)} 
\Bigg ( \frac {1}{e^{({E_i}^* (k) -{\mu_i}^*)/T}+1}
+ \frac {1}{e^{({E_i}^* (k) +{\mu_i}^*)/T}+1} \Bigg ),
\label{scaldens}
\end{align}
where, ${E_i}^*(k)=(k^2+{{m_i}^*}^2)^{1/2}$ and ${\mu _i}^* 
=\mu_i -g_{\omega i}\omega -g_{\rho i}\rho -g_{\phi i}\phi$, are the single 
particle energy and the effective chemical potential
for the baryon of species $i$, and
$\gamma_i$=2 is the spin degeneracy factor. Also, $m_i^*$ = $- g_{\sigma i} \sigma - g_{\zeta i} \zeta - g_{\delta i}\delta$ is the effective mass of the baryons in the asymmetric hadronic medium. Parameters $g_{\sigma i}$, $g_{\zeta i}$ and $g_{\delta i}$ are fitted to reproduce the vacuum baryon masses \cite{papa_nuclei}. In \cref{chi} $\sigma_0$, $\zeta_0$ and $\chi_0$ denote the vacuum values of the scalar fields $\sigma$, $\zeta$ and $\chi$. The parameter $d$ is a constant having value $2/11$, determined through the QCD beta function at one loop level for three colors $N_c$ and three flavors $N_f$ \cite{papa_nuclei}. 
For given density $\rho_B$ of the baryonic medium, we solve  coupled equations of motion of scalar fields using mean field approximation for the different values of strangeness fractions $f_s$, isospin asymmetric parameter $I$ and temperature $T$. The strangeness fraction is defined as $f_s$ = $\frac{\Sigma_i |s_i|\rho_i}{\rho_B}$, here $s_i$ is the number of strange quarks and $\rho_i$ is number density of $i^{th}$ baryon defined by $\rho_{i} = \gamma_{i}\int\frac{d^{3}k}{(2\pi)^{3}} 
 ( \frac {1}{e^{({E_i}^* (k) -{\mu_i}^*)/T}+1}
+ \frac {1}{e^{({E_i}^* (k) +{\mu_i}^*)/T}+1})$. Further, the isospin asymmetric parameter is defined as $I = -\frac{\Sigma_i I_{3i} \rho_i}{2\rho_B}$, where $I_{3i}$ is the $z$-component of  the isospin for the $i^{\text{th}}$ baryon \cite{arvind2}. 
 Further, to evaluate the shift in masses and decay constants of $D_s$ and $B_s$ mesons using QCD sum rule analysis, we shall need strange quark condensate $\left\langle \bar{s}s\right\rangle$. To find this, we  use the explicit chiral symmetry breaking term to represent the strange quark condensate in terms of the strange scalar field $\zeta$ as \cite{ss},
\begin{align}
\left\langle \bar{s}s\right\rangle
= \frac{1}{m_{s}}\left( \frac{\chi}{\chi_0} \right)^2\left( \sqrt{2}m_K^2 f_K  - \frac{1}{\sqrt{2}} m_ {\pi}^2 f_{\pi}\right) \zeta,
\label{qd}
\end{align}
here $m_s$ denotes the mass of strange quark. Also, $\chi_0$ denotes the vacuum value of the dilaton field $\chi$.  
To express gluon condensate in terms of above calculated fields, we obtain the energy  momentum tensor through the scale  breaking term of the effective Lagrangian density. In the limit of finite quark masses, we equate the trace of energy momentum tensor calculated in effective chiral SU(3) model with the trace of energy momentum tensor calculated in QCD, which is actually the gluon condensates \cite{papa_nuclei,arvind2} given as,

\begin{align}
\left\langle  \frac{\alpha_{s}}{\pi} {G^a}_{\mu\nu} {G^a}^{\mu\nu}
\right\rangle =  \frac{8}{9} \Bigg [(1 - d) \chi^{4}
+\left( \frac {\chi}{\chi_{0}}\right)^{2}
\left( m_{\pi}^{2} f_{\pi} \sigma
+ \big( \sqrt {2} m_{K}^{2}f_{K} - \frac {1}{\sqrt {2}}
m_{\pi}^{2} f_{\pi} \big) \zeta \right) \Bigg ].
\label{glu}
\end{align}
This procedure helps to calculate the medium modified gluon condensates through the medium modified $\sigma$, $\zeta$ and $\chi$ fields. In the above equation, $d$ is  constant with value $(2/11)$ and it is calculated through the one loop beta function for three flavors and colors of QCD \cite{papa_nuclei}.

  \section{QCD sum rule for $D_s$ and $B_s$ mesons}
  \label{QCD}
We will use the output of the chiral SU(3) model (medium modified  strange and gluon condensates), as an input in the QCD sum rules to investigate the in-medium masses and decay constants of $D_s$ and $B_s$ mesons. QCD sum is an useful technique which relate the phenomenological spectral parameters with the basic properties of QCD. In QCD sum rules we start with two point correlation function
 \begin{align}
\Pi (q) = i\int d^{4}x\ e^{iq_{\mu} x^{\mu}} \langle \mathcal{T}\left\{J_5(x)J_5^{\dag}(0)\right\} \rangle_{\rho_B, T} ,
\label{tw}
\end{align}
where $\mathcal{T}$ is the time-ordered covariant operator acting on  pseudo-scalar currents for the  $D_s$  meson, represented as \cite{haya2004, lucha}
\begin{align}
 J_5(x) &= J_5^\dag(x) = (m_c+m_s)\frac{\bar{c}(x)i\gamma_5 s(x)+{c}(x)i\gamma_5 \bar{s}(x)}{2} .
 \label{psc}
 \end{align}
 For $B_s$ meson $c(x)$ quark field will be replaced by $b(x)$ quark field. Here, by using the above mentioned psuedoscalar current we will concentrate on the averaged shift in the masses and decay constants of $D_s$ and $\bar{D_s}$ (Similarly for $B_s$ and $\bar{B_s}$) mesons. On the other hand, in \cite{hilger2009}, authors observed the splitting between $D_s$ and $\bar{D_s}$ mesons by separating two point correlation function into even and odd part. Further, the pseudoscalar current is related to the decay constant as $\langle 0|J_5|D(k) \rangle = f_D m_D^2/(m_c+m_s) $, where $D(k)$ is the state of $D$ meson in four momentum $k$, $m_D$ and $f_D$ are the masses and decay constants of $D$ meson, respectively \cite{haya2004, lucha}. In  literature, using heavy quark limit, the mass of light quark is neglected as compared to the charm quark mass \cite{numb, haya2000}. Later on, we will point out that, inclusion of the mass of strange quark will have negligible effect on the results of this paper. Therefore, in the present calculation, we will work in heavy quark limit, i.e.,  $m_c+m_s \simeq m_c$.  Moreover, using Fermi gas approximation, in the rest frame of nucleons, we decompose the two point correlation function into vacuum part, nucleon dependent part and pion bath term, i.e.,
 \begin{align}
\Pi (q) =\Pi_{0} (q)+ \frac{\rho_B}{2m_N}T_{N} (q) + \Pi_{P.B.}(q,T)\,,
\label{pb}
 \end{align}

 where $T_N (q)$ is the forward scattering amplitude, $\rho_B$ and $m_N$ denote the total baryon density and nucleon mass, respectively. The third term represents the thermal correlation function and is defined as \cite{Elet}
\begin{align} \label{pb2}
\Pi_{P.B.}(q, T) = i\int d^{4}x\ e^{iq_{\mu} x^{\mu}} \langle \mathcal{T}\left\{J_5(x)J_5^\dag(0)\right\} \rangle_{T},
\end{align}
where $\langle \mathcal{T}\left\{J_5(x)J_5^\dag(0)\right\}\rangle_{T}$  is the thermal average of the time ordered product of the pseudoscalar currents.
The thermal average of any operator $\mathcal{O}$ is given by \cite{Elet}
\begin{align}
\left\langle \mathcal{O} \right\rangle _T = \frac{Tr \left\lbrace  \text{exp}\left(-H/T\right) \mathcal{O}\right\rbrace}{Tr \left\lbrace \text{exp}\left(-H/T\right)\right\rbrace}.
\end{align}
In above $Tr$ denotes the trace over complete set of states and $H$ is the QCD Hamiltonian. The factor $\frac{\text{exp}\left(-H/T\right)}{Tr \left\lbrace\text{exp}\left(-H/T\right) \right\rbrace}$  is the thermal density matrix of QCD. In \cref{pb}, the third term corresponds to the pion bath term and had been widely used in the past to consider the effect of temperature of the medium \cite{zsch,kwon}. 
  We shall consider the effect of temperature through the temperature dependence of the condensates, calculated in the chiral SU(3) model and thus, we neglect the third term in \cref{pb}.  Further, to calculate the shift in the masses and the decay constants,  we express the scattering amplitude $T_N (q)$, near the pole position of the pseudoscalar mesons in terms of the spectral density \cite{koi}. This spectral density in the limit of $q$ $\to$ 0 can be further parameterized in terms of three unknown parameters $a$, $b$ and $c$, given as    \cite{wang1,wang2,haya2000},
\begin{align} 
\rho(\omega,0) &= -\frac{f_{D_s/B_s}^2m_{D_s/B_s}^4}{\pi m_{c/b}^2} \nonumber
 \mbox{Im} \left[\frac{{{T}_{D_s/B_s}}(\omega,{\bf 0})}{(\omega^{2}-
m_{D_s/B_s}^2+i\varepsilon)^{2}} \right] + \cdots  \\ \nonumber
&=-\frac{f_{D_s/B_s}^2m_{D_s/B_s}^4}{\pi
m_{c/b}^2}\Big\{\mathbf{Im}\frac{1}{(\omega^2-m_{D_s/B_s}^2+i\varepsilon)^2}
\mathbf{Re}[\mathbf{T}_{D_s/B_s}(\omega,0)]
\nonumber \\
&+ \mathbf{Re}\frac{1}{(\omega^2-m_{D_s/B_s}^2+i\varepsilon)^2}
\mathbf{Im}[\mathbf{T}_{D_s/B_s}(\omega,0)]\Big\}+...
\label{a1}
\end{align}
\begin{align}
& = a\,\frac{d}{d\omega^2}\delta(\omega^{2}-m_{D_s/B_s}^2)
 +
b\,\delta(\omega^{2}-m_{D_s/B_s}^2) + c\,\theta(\omega^{2}-s_{0})\,.
\label{a2}
\end{align}

Here,  $m_{D_s/B_s}$ and $f_{D_s/B_s}$ are the masses and decay constants of $D_s/B_s$ mesons. In the above equation, first term represents the double pole term and corresponds to the on shell effect of the $T$-matrix. The second term is the single pole term and denotes the off-shell effect of the $T$-matrix.  
 The third term proportional to $c$, corresponds to the continuum   term and the  contribution of higher states. Here the possibility of the errors may occur only through the third term  of \cref{a2} and therefore special care is taken while dealing with it \cite{azizi}.In this respect, the value of continuum threshold parameter $s_0$, is fixed so as to reproduce the vacuum masses for $D_s$ and $B_s$ mesons. Further, using \cref{a2,pb,pb2} the shift in masses and decay constants of $D_s$/$B_s$ mesons can be expressed as in terms of $a$ and $b$ as \cite{azizi,wang2}

    \begin{equation}
\delta m_{D_s/B_s} = 2\pi \frac{m_N + m_{D_s/B_s}}{m_N m_{D_s/B_s}} \rho_N a_{D_s/B_s},
\label{masshift}
\end{equation}
  and 
 \begin{equation}
 \delta f_{D_s/B_s} =  \frac{m_{c/b} ^2}{2f_{D_s/B_s} m^4}\left(\frac{b \rho_N}{2m_N} - \frac{4 f_{D_s/B_s}^2 m_{D_s/B_s}^3 \delta m_{D_s/B_s}}{m_{c/b} ^2}\right).
 \label{decayshift}
\end{equation} 
Clearly, to evaluate the shift in mass and decay constant of $D_s/B_s$ mesons we need to calculate the unknown parameters $a$ and $b$. 
 
 Therefore, in this respect we apply the Borel transformation on the forward scatterring amplitude $T_N(\omega,0)$ in the hadronic side using the assumption of quark hadron duality. Also, we apply the Borel transformation on the forward scattering amplitude $T_N(\omega,0)$ calculated in operator product expansion (OPE) side in the rest frame of the nuclear matter. Then we equate these Borel transformed hadronic and OPE side of the $T_N(\omega,0)$ functions, and further this analysis lead us to the relation between  $a$, $b$ and the strange quark and gluon condensates as in \cite{azizi,wang2}. Thus, we have

\begin{align}
a E_1 + bE_2 = E_3 
\label{ea}
\end{align} 

where,

\begin{align}
E_1 = & \left\{\frac{1}{M^2}\exp\left(-\frac{m_{{D_s/B_s}}^2}{M^2}\right) - \frac{s_0}{m_{{D_s/B_s}}^4} \exp\left(-\frac{s_0}{M^2}\right)\right\} \nonumber,
\\
E_2 = & \left\{\exp\left(-\frac{m_{{D_s/B_s}}^2}{M^2}\right) - \frac{s_0}{m_{{D_s/B_s}}^2} \exp\left(-\frac{s_0}{M^2}\right)\right\}\nonumber,
\end{align}
and,
\begin{align}\label{qcds}
E_3 = &  \frac{2m_N(m_H+m_N)}{(m_H+m_N)^2-m_{{D_s/B_s}}^2}\left(\frac{f_{{D_s/B_s}}m_{{D_s/B_s}}^2g_{{D_s/B_s}NH}}{m_{c/b}}\right)^2\nonumber\\
\times & \left\{ \left[\frac{1}{M^2}-\frac{1}{m_{{D_s/B_s}}^2-(m_H+m_N)^2}\right] 
\exp\left(-\frac{m_{{D_s/B_s}}^2}{M^2}\right)\right.\nonumber\\
+&\left.\frac{1}{(m_H+m_N)^2-m_{{D_s/B_s}}^2}\exp\left(-\frac{(m_H+m_N)^2}{M^2}\right)\right\}\nonumber\\
=&-\frac{m_{c/b}\langle\bar{s}s\rangle_N}{2}\left\{1+\frac{\alpha_s}{\pi} \left[ 6-\frac{4m_{c/b}^2}{3M^2} \right.\right.
\left.\left.-\frac{2}{3}\left( 1-\frac{m_{c/b}^2}{M^2}\right)\log\frac{m_{c/b}^2}{\mu^2}-2\Gamma\left(0,\frac{m_{c/b}^2}{M^2}\right)\exp\left( \frac{m_{c/b}^2}{M^2}\right) \right]\right\} \nonumber\\
\times &\exp\left(- \frac{m_{c/b}^2}{M^2}\right) \nonumber\\
+&\frac{1}{2}\left\{-2\left(1-\frac{m_{c/b}^2}{M^2}\right)\langle s^\dag i D_0s\rangle_N +\frac{4m_{c/b}
}{M^2}\left(1-\frac{m_{c/b}^2}{2M^2}\right)\langle \bar{s} i D_0 i D_0s\rangle_N+\frac{1}{12}\langle\frac{\alpha_sGG}{\pi}\rangle_N\right\} \nonumber\\
\times &\exp\left(- \frac{m_{c/b}^2}{M^2}\right).
\end{align}
We differentiate \cref{ea} with respect to  $\frac{1}{M^2}$ to find another equation

\begin{align}
a {E'_1} + bE'_2 = E'_3 \nonumber\\
\end{align}\label{Eaa}
where 
$E_i\textquotesingle$ denotes the first derivative. Finally, $a$ and $b$ will be calculated as, $a$ = $\frac{E_3 E_2 - E'_3 E_2}{E_1 E'_2- E'_1 E_2}$ and $b$ = $\frac{E_3 E'_1 - E_3 E_1}{E_2 E'_1- E'_2 E_1}$.

Furthermore, the nucleon expectation values of the various condensates appearing in  \cref{qcds}, can be calculated as \cite{hilger2009}  
\begin{equation}
\mathcal{O}_{N} = \left[ \mathcal{O}_{\rho_{B}}  - \mathcal{O}_{vacuum}\right] \frac{2m_N}{\rho_B},
\label{condexp}
\end{equation}
where, $\mathcal{O}_{\rho_{B}}$ and $\mathcal{O}_{vacuum}$ denote the expectation values of the operators at finite baryonic density and vacuum, respectively. Explicitly, the nucleon expectation values of strange quark and gluon condensates can be written as 
\begin{equation}
\left\langle \bar{s}s \right\rangle_{N} = \left[  \left\langle s\bar{s} \right\rangle _{\rho_{B}}  - \left\langle  \bar{s}s \right\rangle_{vacuum}\right] \frac{2m_N}{\rho_B},
\label{ucondexp1}
\end{equation}

and 
\begin{equation}
\left\langle  \frac{\alpha_{s}}{\pi} {G^a}_{\mu\nu} {G^a}^{\mu\nu} 
\right\rangle_{N} = \left[ \left\langle  \frac{\alpha_{s}}{\pi} {G^a}_{\mu\nu} {G^a}^{\mu\nu} \right\rangle_{\rho_B} -  \left\langle  \frac{\alpha_{s}}{\pi} {G^a}_{\mu\nu} {G^a}^{\mu\nu} \right\rangle_{vacuum} \right]\frac{2m_N}{\rho_B},
\label{Gcondexp1}
\end{equation}
respectively.
Also, 
\begin{align}
\langle\bar{s}g_s\sigma Gs\rangle_{\rho_B} = \lambda^{2}\left\langle \bar{s}s \right\rangle_{\rho_{B}} + 3.0 \textrm{GeV}^{2}\rho_{B},
\label{cond2}
\end{align}
and
\begin{align}
\langle \bar{s} i D_0 i D_0s\rangle_{\rho_B} + \frac{1}{8}\langle\bar{s}g_s\sigma Gs\rangle_{\rho_B} =  0.3 \textrm{GeV}^{2}\rho_{B}.
\label{cond3}
\end{align}

 Here we calculate the in-medium values of condensates $ \left\langle  \bar{s}s
\right\rangle_{\rho_B}$,
$\left\langle  \frac{\alpha_{s}}{\pi} {G^a}_{\mu\nu} {G^a}^{\mu\nu}
\right\rangle_{\rho_B}$ using \cref{qd,glu} in chiral SU(3) model. Also, we note that the effect of temperature on the shift in masses and decay constants of $D_s$ and $B_s$ mesons is taken through temperature dependence of the quark and gluon condensates appearing in \cref{qd,glu}, which is further considered through the temperature dependence of the $\sigma, \zeta$ and  $\chi$ fields through \cref{scaldens}.  Further, using these in-medium condensates in \cref{cond2,cond3}, we can calculate in-medium values of $\langle\bar{s}g_s\sigma Gs\rangle_{\rho_B}$ and $\langle \bar{s} i D_0 i D_0s\rangle_{\rho_B}$  respectively. Thus, through these in-medium condensates we can calculate the in-medium masses and decay constants of $D_s$ and $B_s$ mesons.

\section{$^3P_0$ model} 
\label{3P0}
We use $^3 P_0$ model to calculate the effect of shift in mass of $D_s(1968)$ meson on the   parital decay widths of $D_s^*(2715)$ and $D_s^*(2860)$ states decaying to ($D_s , \eta$) channel.  $^3 P_0$ model was firstly invented by Micu \cite{micu}, then developed for the OZI allowed decay of mesons \cite{yao}. In past, $^3 P_0$ model had been widely used for strong decays of hidden charmed states \cite{barn1, barn2}, open charmed bottom states \cite{close, sego} as well as hidden bottom \cite{ferre1, ferre2} and open bottom states \cite{sego, ferre3}. This assumes the creation of quark and anti-quark pair with quantum numbers $0^{++}$. In the present work of two body decay of $D_s^*(2715)$ and $D_s^*(2860)$ mesons, we consider the non-relativistic transition operator \cite{Y(4040)}, and arrive at the helicity amplitude given by \cite{liu} 
\begin{widetext}
\begin{align}\label{M}
 \mathcal{M}^{M_{J_A } M_{J_{B} } M_{J_C }} = \gamma  \sqrt {8E_A E_B E_C } \sum_{\substack{M_{L_A } ,M_{S_A } ,M_{L_B }, \\M_{S_B } ,M_{L_C} ,M_{S_C } ,m} }\langle {1m;1 - m}|{00} \rangle \nonumber \\
 \times \langle {L_A M_{L_A } S_A M_{S_A } }| {J_A M_{J_A } }\rangle \langle L_B M_{L_B } S_B M_{S_B }|J_B M_{J_B } \rangle\langle L_C M_{L_C } S_C M_{S_C }|J_C M_{J_C }\rangle \nonumber \\
  \times\langle\varphi _B^{13} \varphi _C^{24}|\varphi _A^{12}\varphi _0^{34} \rangle
\langle \chi _{S_B M_{S_B }}^{13} \chi _{S_C M_{S_C } }^{24}|\chi _{S_{A} M_{S_{A} } }^{12} \chi _{1 - m}^{34}\rangle I_{M_{L_{B} } ,M_{L_C } }^{M_{L_{C}} ,m} (\textbf{K}).
\end{align}
\end{widetext}
In above,  $E_{A}$= $m_{A}$, $E_{B}$ = $\sqrt{m_{B}^{*2} + K_B^2}$ and $E_{C}$ = $\sqrt{m_{C}^{*2} + K_C^2}$ represent the energies of respective mesons. Here $m^*_B$ and $m^*_C$ are the in-medium masses of $D_s(1968)$ and $\eta$ mesons respectively.
We follow the literature  \cite{Y(4040), liu}, and calculate the spin matrix elements $\langle  \chi _{S_B M_{S_B }}^{13} \chi _{S_C M_{S_C } }^{24}|\chi _{S_{A} M_{S_{A} } }^{12} \chi _{1 - m}^{34}\rangle$ in terms of the Wigner's 9j symbol, whereas the flavor matrix element $\langle\varphi _B^{13} \varphi _C^{24}|\varphi _{A}^{12}\varphi _0^{34} \rangle$ is expressed in terms of isospin of quarks \cite{yao}. Further, $I_{M_{L_{B} } ,M_{L_C } }^{M_{L_{C}} ,m} (\textbf{K})$ is the spatial integral for the general decay ($A$ $\to$ $B C$) and is written in terms of Fourier transformed harmonic oscillator meson wave function, i.e, 

\begin{align}\label{I}
I_{M_{L_B } ,M_{L_C } }^{M_{L_{A} } ,m} (\textbf{K}) = \int d \textbf{k}_1 d \textbf{k}_2 d \textbf{k}_3 d \textbf{k}_4 \delta ^3 (\textbf{k}_1 + \textbf{k}_2)\delta ^3 (\textbf{k}_3+ \textbf{k}_4)\delta ^3 (\textbf{k}_B- \textbf{k}_1- \textbf{k}_3 )\nonumber \\
\times \delta ^3 (\textbf{k}_C - \textbf{k}_2 -\textbf{k}_4) \Psi _{n_B L_B M_{L_B } }^* (\textbf{k}_1 ,\textbf{k}_3)\Psi _{n_C L_C  M_{L_C}}^* (\textbf{k}_2 ,\textbf{k}_4)\nonumber \\ \times \Psi _{n_{A} L_{A} M_{L_{A}}} (k_1 ,k_2 )Y _{1m}\left(\frac{\textbf{k}_3-\textbf{k}_4}{2}\right),
\end{align}

where, ${\bf k}_i$ represents the momentum of corresponding quark, $\mathcal{Y}_{1}^{m}(\mathbf{k})\equiv
|\mathbf{k}|^{l}Y_{l}^{m}(\theta_{k},\phi_{k})$ is the solid
harmonic polynomial  of the  quark antiquark pair created in $^3 P_0$ model. 
By taking all these calculations in hand, we use the Jacob-Wick formula to transform the helicity amplitude into partial wave amplitude as follows

\begin{eqnarray}
\mathcal{M}^{JL} (A \to B C) &=& \frac{{\sqrt {2{L} + 1} }}{{2{ J_{A}} + 1}}\sum_{M_{J_B},M_{ J_C}} \langle{{ L}0{J} M_{J_{A}}} |{J_{A} M_{J_{A} } }\rangle \nonumber \\
&&\times \left\langle {{ J_{B}} M_{J_B} { J_C} M_{J_C}}\right|\left. {{J} M_{J_{A} } } \right\rangle M^{M_{J_{A} } M_{J_{B}} M_{J_C } } (\textbf{K}).
\end{eqnarray}

In above, $M_{J_{A}}=M_{J_B}+M_{J_C}$, $|{{J_B}}-{{J_C}}| \leq $ ${{J}}$  $\leq$ $|{{J_B}} +{ J_C}|$  and $|{J} - {L}| \leq { J_{A}} $  $\leq |{ J} + { L}|$. In the present investigation, we chose the harmonic type oscillator wave function for $D_s^*(2860)$ ($n$=1, $L$=2) and $D_s^*(2715)$ ($n$=2, $L$=0) states  as \cite{liu}
\begin{align} \label{D2860}
 \psi({{\mathbf{k}_1,\mathbf{k}_2}}) = \frac{R^{7/2}}{\sqrt{15}\pi^{1/4}}
\mathcal{Y}_{2}^{m}\big(\frac{\mathbf{k}_{1}-\mathbf{k}_{2}}{2}\big)
\exp\Big{[}-{1\over 8}({{\mathbf{k}_1-\mathbf{k}_2})}^2R^2\Big{]},
\end{align}
and
\begin{align} \label{D2715}
\psi({{\mathbf{k}_1,\mathbf{k}_2}}) =
\frac{1}{\sqrt{4\pi}}\bigg (\frac{4R^3}{\sqrt{\pi}}\bigg)^{{1}/{2}}
\sqrt{\frac{2}{3}} \bigg[\frac{3}{2} - \frac{R^2}{4}
 ({{\mathbf{k}_1-\mathbf{k}_2})}^2 \bigg]\exp\Big{[}-{1\over 8}({{\mathbf{k}_1-\mathbf{k}_2})}^2
 R^2\Big{]}.
 \nonumber
\end{align}
Also, the harmonic oscillator wave function for the daughter mesons in ground state ($n$=1, $L$=0) will be
\begin{align}
\psi({\bf k}_1,{\bf k}_2) = \bigg (\frac{R^{2}}{\pi}\bigg)^{3/4}\exp \left( -\frac{({\bf k}_1-{\bf k}_2)^2}{8}R^2 \right).
\end{align}

In the above equations $R$ denote the radius of respective meson. We finally calculate the decay width, using
\begin{align}
\Gamma  = \pi^2 \frac{|{\textbf{K}}|}{m_{A}^2}\sum_{JL} |{\mathcal{M}^{JL}}|^2,
\label{G}
\end{align}
where $|\textbf{K}|$ is the momentum of the $B$ and $C$ mesons in the rest mass frame of A meson and is given by, 
\begin{align}
 |\textbf{K}|= \frac{{\sqrt {[m_{A}^2  - (m^*_B  - m^*_C )^2 ][m_{A}^2  - (m^*_B  + m^*_C )^2 ]} }}{{2m_{A} }}.
\label{K}
\end{align}

 Thus, through the in-medium mass of $D_s(1968)$ and $\eta$ mesons, the in-medium partial decay widths of $D_s^*(2715)$ and $D_s^*(2860)$ states decaying to ($D_s(1968)$, $\eta$) can be calculated.

\section{Results and discussion}
\label{results}
In this section, we shall discuss the various results of the present investigation. We use parameters, nuclear saturation density, $\rho_0$ = 0.15 fm$^{-3}$, the average values of coupling constants $g_{{{D_s}N\Lambda_c}}$ $\approx$ $g_{{{D_s}N\Sigma_c}}$ $\approx$ $g_{{{B_s}N\Lambda_c}}$ $\approx$ $g_{{{B_s}N\Sigma_c}}$ $\approx$ 6.74, the values of continuum threshold parameter $s_0$, for $D_s$ and $B_s$ mesons as 7.3 and 36 GeV$^2$, respectively.  We take the vacuum masses and decay constants of $D_s$($B_s$) mesons as, 1.968(5.37) and 0.240(0.231) GeV, respectively \cite{review}. Moreover, the shift in masses and decay constants of $D_s$ and $B_s$ mesons are represented as a function of squared Borel mass parameter, $M^2$.  We choose the proper Borel window where we find the least variation in the masses and decay constants. For example, we chose the Borel windows for masses of $D_s$ and $B_s$ meson as (4-6) and (28-32) GeV$^2$, respectively and, for decay constants we chose the respective Borel windows as (2-4) and (28-32) GeV$^2$. 
The results of the present work are much sensitive to the  choice of the Borel window \cite{haya2000,wang2,wang1,azizi}. 
Because of this we take different Borel windows for the masses and decay constants of $D_s$  meson. 
 On the other hand, for  $B_s$ meson we observed same stable region for the shift in mass and decay constant.

 \begin{table}
 \begin{tabular}{|l|l|l|l|l|l|l|l|l|l|}
\hline
 & & \multicolumn{4}{c|}{I=0}    & \multicolumn{4}{c|}{I=0.5}   \\
\cline{3-10}
&$f_s$ & \multicolumn{2}{c|}{T=0} & \multicolumn{2}{c|}{T=100MeV}& \multicolumn{2}{c|}{T=0}& \multicolumn{2}{c|}{T=100MeV}\\
\cline{3-10}
&  &$\rho_0$&$4\rho_0$ &$\rho_0$  &$4\rho_0$ & $\rho_0$ &$4\rho_0$&$\rho_0$&$4\rho_0$ \\ \hline 
$\delta m_{D_s}$ & 0& -41&-65 &-36&-60&-40&-61&-36 &-59\\ \cline{2-10}
&0.5&-68  &-128  & -59 & -114 &-73 &-141 & -63 & -121 \\ \cline{1-10}
$\delta m_{B_s}$&0&-285 & -515 & -251 &-488 &-280 &-493 & -251 &-477 \\  \cline{2-10}
&0.5&-451 &-909 &-398 &-821 & -482 &-989 & -420&-864 \\  \hline
$\delta {f_{D_s}}$& 0&-4&-4.5 &-3.4&-4&-3.9&-4.1&-3.4 &-3.8\\ \cline{2-10}
&0.5&-7.2  &-12  &-6.2 & -10 &-7 &-13 & -6.5 & -11 \\ \cline{1-10}
$\delta {f_{B_s}}$&0&-39 & -71 & -35 &-67.8 &-39 &-68 & -35 &-66 \\  \cline{2-10}
&0.5&-63 &-127 &-55 &-114 & -67 &-138 & -58.8&-121 \\  \hline
\end{tabular}
\caption{In above, we tabulate the values of shift in masses and decay constants  of $D_s$ and $B_s$ mesons (in units of MeV).}
\label{table:tbl}
\end{table}

\subsection{Shift in masses and decay constants}
\label{sub_mass}   
\begin{figure}
\centering
\includegraphics[width=14cm,,height=12cm]{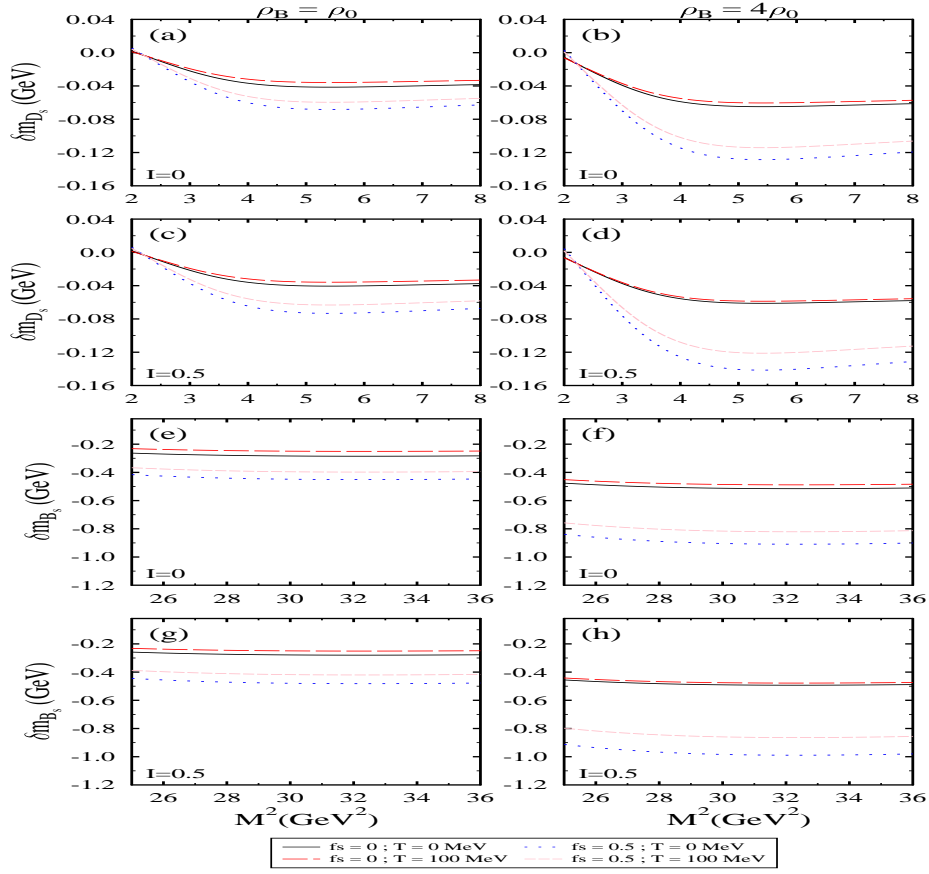}
\caption{Figure shows the variation of shift in mass of pseudoscalar $D_s$ and $B_s$ mesons as a function of squared Borel mass parameter, $M^2$ for baryonic densities $\rho_0$ and $4\rho_0$. The results are given for the isospin asymmetric parameters $I = 0$ and $0.5$, temperatures $T = 0$ and 100 MeV and strangeness fractions $f_s = 0$ and $0.5$.}\label{fig:mass}
\end{figure}

\begin{figure}
\centering
\includegraphics[width=14cm,,height=12cm]{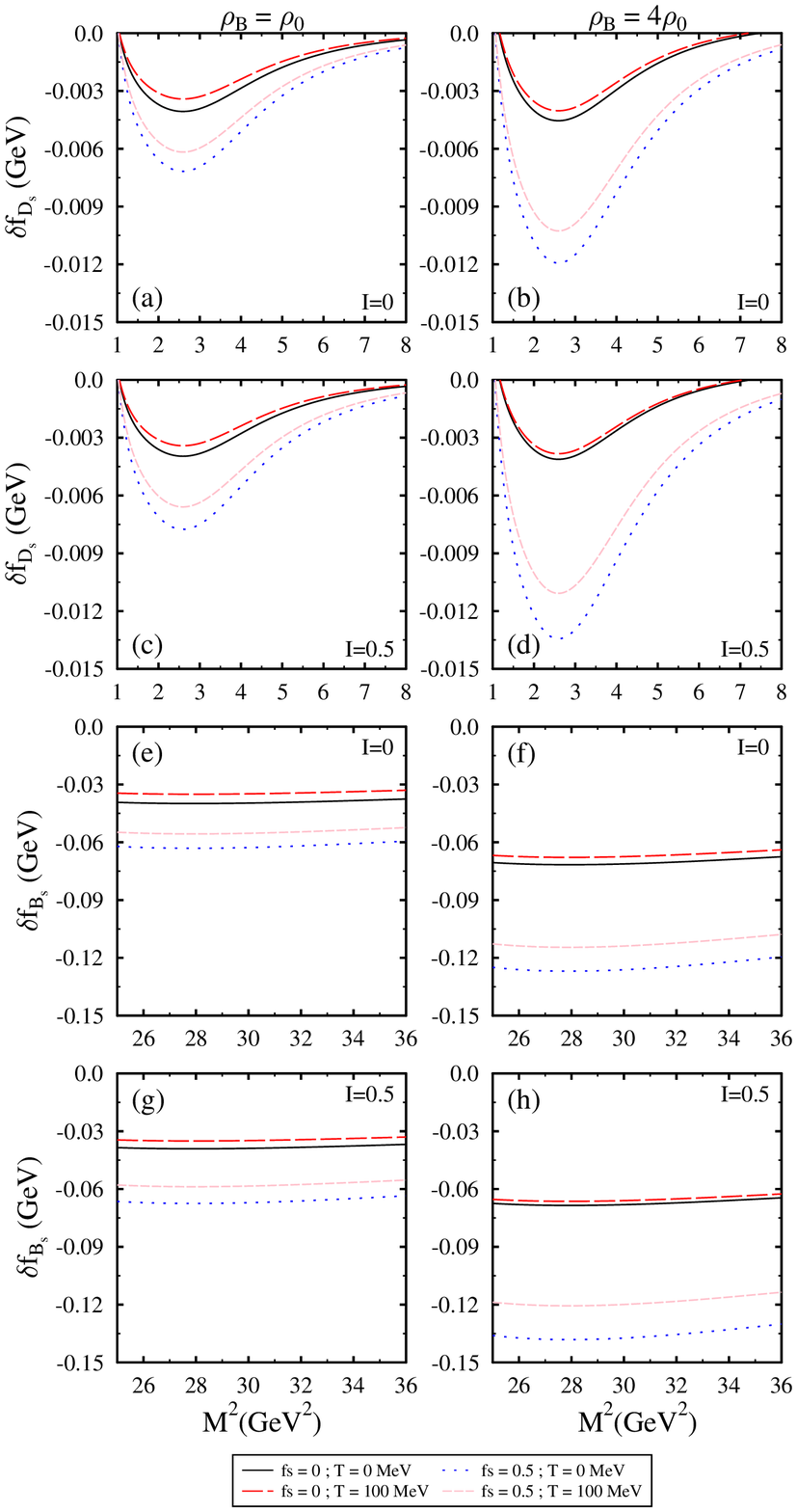}
\caption{Figure shows the variation of shift in decay constants of pseudoscalar $D_s$ and $B_s$ mesons as a function of squared Borel mass parameter, $M^2$ for baryonic densities $\rho_0$ and $4\rho_0$. The results are given for isospin asymmetric parameters $I = 0$ and $0.5$, temperatures $T = 0$ and 100 MeV and strangeness fractions $f_s = 0$ and $0.5$.}\label{fig:decay}
\end{figure}

  In \cref{fig:mass} (\cref{fig:decay}) we represent the shift in masses (decay constants) of $D_s$ and $B_s$ mesons in symmetric as well as in asymmetric hot and dense hadronic medium as a function of squared Borel mass parameter, $M^2$. In \cref{table:tbl}, we represent the shift in masses and decay constants of these mesons for the baryonic densities, $\rho_0$ and 4$\rho_0$, strangeness fractions, $f_s$=0 and $f_s$=0.5, temperatures, $T$=0 and $T$=100 MeV and  isospin asymmetric parameters, $I$=0 and $I$=0.5. For any particular value of temperature of the symmetric medium, we notice a drop (from vacuum values) in the masses and decay constants of $D_s$ and $B_s$ mesons as a function of either baryonic density, $\rho_B$,  or strangeness fractions, $f_s$. For example, in symmetric nuclear medium at nuclear saturation density, $\rho_0$, and temperature, $T$=0, the values of masses and decay constants of $D_s$ ($B_s$) mesons decrease  by 2$\%$ (5$\%$), and 1.6$\%$ (16$\%$), respectively from their vacuum values. Likewise, at baryonic density, 4$\rho_0$, the values of percentage drop in masses and decay constants change to 3.3$\%$ (9.6$\%$) and  1.8$\%$ (30$\%$), respectively. On the other hand, at baryon density, $\rho_0$  
 and temperature $T$=0, if we move from symmetric nuclear medium ($f_s$=0) to symmetric strange hadronic medium ($f_s=0.5$), then the percentage drop in the masses and decay constants of $D_s$ ($B_s$) mesons observed to be 3.4$\%$ (8.4$\%$) and 3$\%$ (27$\%$), respectively from their vacuum values.
Likewise, at high baryonic density, 4$\rho_0$,  the magnitude of drop enhance to 6.5$\%$ (16$\%$) and 5$\%$ (55$\%$), respectively.
 
This drop of masses and decay constants of $D_s$ and $B_s$ mesons in the hadronic medium can be understood in terms of in-medium dependence of the quark and gluon condensates appearing in \cref{qcds}. As can be seen from \cref{qd,glu},  the scalar strange condensates $\left\langle \bar{s}s\right\rangle$, is proportional to $\zeta$ field, and the scalar gluon condensates $\left\langle  \frac{\alpha_{s}}{\pi} {G^a}_{\mu\nu} {G^a}^{\mu\nu}\right\rangle$, depend upon $\sigma$, $\zeta$ and $\chi$ fields. Therefore, the behaviour of scalar fields $\sigma$, $\zeta$ and $\chi$, in the hadronic medium is reflected through the in-medium strange quark and gluon condensates and this is further reflected through the in-medium masses and decay constants of $D_s/B_s$ mesons.  As the $\zeta$ meson contain strange quark content $s$, therefore this is much sensitive to the presence of hyperons in the medium. Also, we notice that the magnitude of $\zeta$ field decreases as a function of baryonic density and strangeness fraction of the medium. Also, the dilaton field $\chi$ is observed to have least variation in hadronic matter as compared to the scalar fields $\sigma$ and $\zeta$
\cite{Rahul}. For example, at finite baryonic density, 4$\rho_B$ and in zero temperature symmetric nuclear matter, the value of $\sigma$, $\zeta$ and $\chi$ fields decrease by 68$\%$, 14$\%$ and 2.9$\%$, respectively, as compared to their vacuum values. Likewise, the corresponding values of  $\left\langle \bar{s}s\right\rangle$ and $\left\langle  \frac{\alpha_{s}}{\pi} {G^a}_{\mu\nu} {G^a}^{\mu\nu}\right\rangle$ decrease by 20$\%$ and 11$\%$. 
On the other hand, the inclusion of hyperons along with the nucleons ($f_s$=0.5) in the symmetric hadronic matter, cause further decrease in the magnitude of $\sigma$, $\zeta$ and $\chi$ field by 68.5$\%$, 26$\%$ and 1$\%$, respectively, from their vacuum values.  Likewise, the magnitude of $\left\langle \bar{s}s\right\rangle$ and $\left\langle  \frac{\alpha_{s}}{\pi} {G^a}_{\mu\nu} {G^a}^{\mu\nu}\right\rangle$ decrease by 32$\%$ and 11$\%$, respectively from its vacuum values. 
Therefore, this decrease in the values of the strange quark condensates and gluon condensates cause  drop in the masses and decay constants of $D_s/B_s$ mesons. 

However, we observe the opposite effect of the temperature, i.e., keeping the other parameters of medium constant, the masses and decay constants of $D_s$ and $B_s$ mesons increase as a function of temperature of the medium. For example, at baryonic density, 4$\rho_0$, temperature, $T$=100 MeV and in symmetric nuclear matter, the percentage drop in the mass and decay constant of $D_s$ ($B_s$) mesons observed to be 3$\%$ (9$\%$) and 1.6$\%$ (29$\%$), respectively. Clearly, these values are slight lower than the zero temperature situations which has been mentioned earlier.  
 The above behaviour is caused by competing effect of the thermal distribution functions and the contributions from higher-momentum states which further cause increase in the value of  $\zeta$ field at finite temperature of the medium, as discussed in detail in ref. \cite{Rahul}.   

 \begin{figure}
\centering
\includegraphics[width=17cm,,height=10cm]{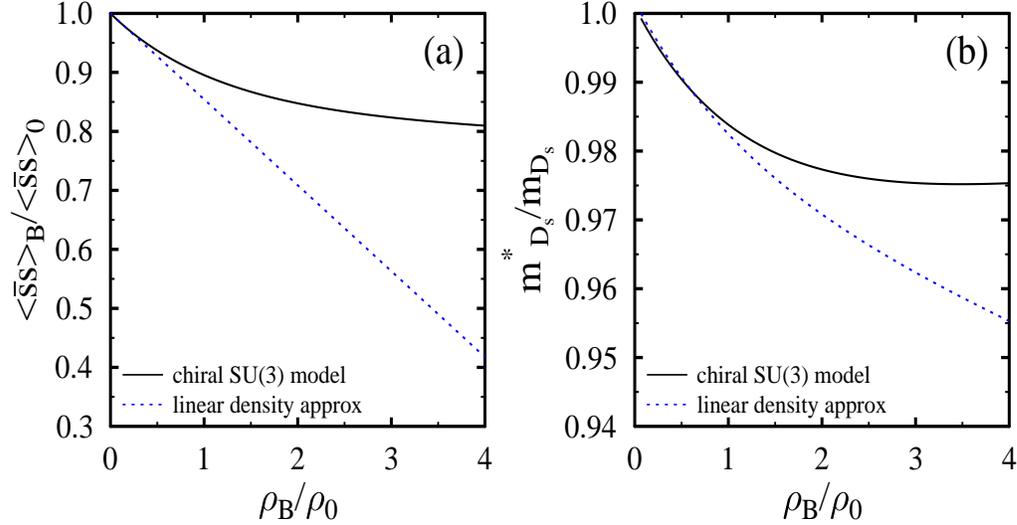}
\caption{We represent the variation of strange quark condensates (subplot (a)) and mass of $D_s(1968)$ meson(subplot(b)), as a function of baryonic density, calculated in linear density approximate QCD sum rules and compared with chiral SU(3) model, in cold symmetric nuclear medium.}\label{fig:valid}
\end{figure}

\begin{figure}
\centering
\includegraphics[width=17cm,,height=10cm]{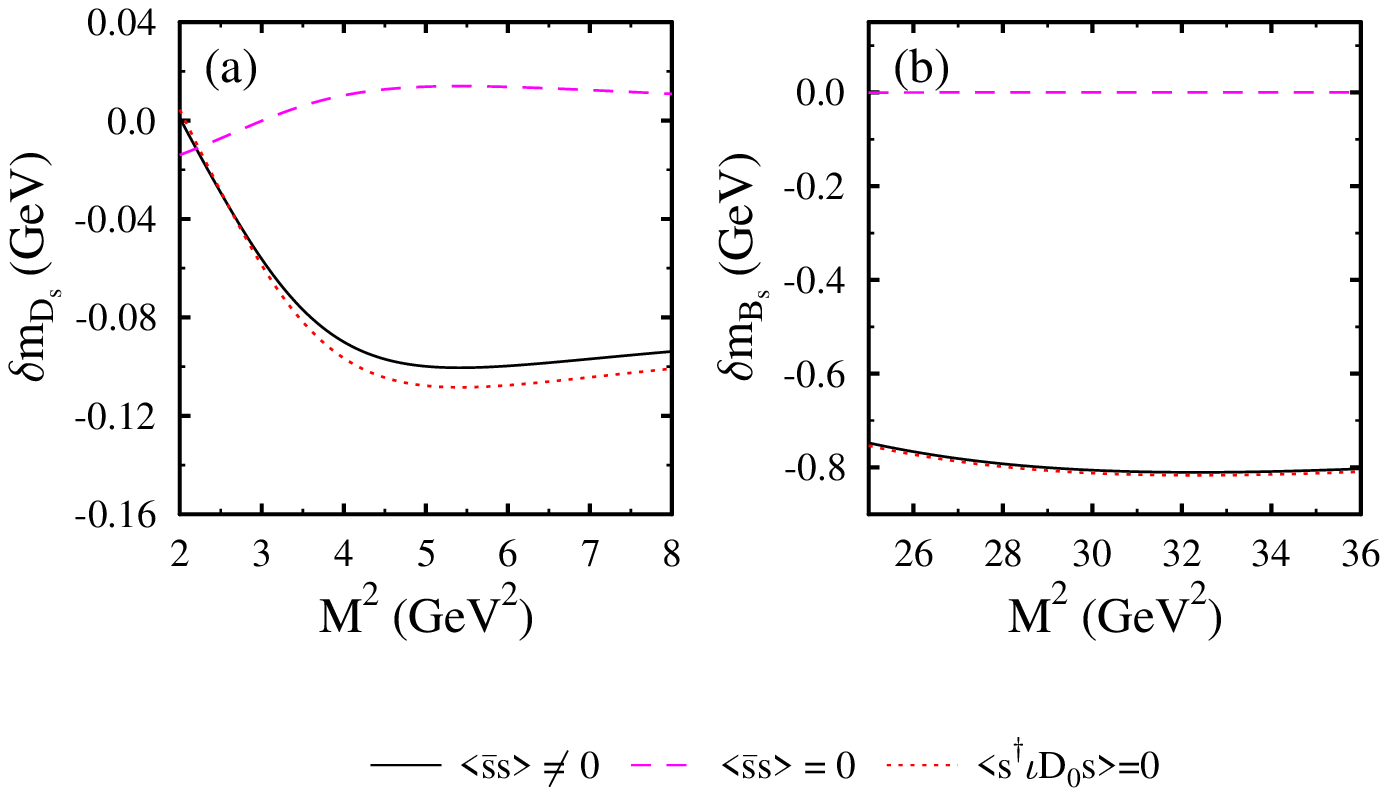}
\caption{Figure shows the contribution of individual condensates to the shift in masses of $D_s$ and $B_s$ mesons, in symmetric hyperonic ($f_s=0.5$) medium, at temperature $T=100$ MeV, baryonic density $\rho_B=4\rho_0$.}\label{fig:individual}
\end{figure}

Furthermore, we observe that the shift in masses and decay constants of $D_s$ ($B_s$) meson is not much sensitive to the isospin asymmetric properties of the medium. As mentioned earlier,  $\zeta$ contain $s$ quark, therefore it does not change appreciably as a function of  isospin asymmetry  of the medium. As a result, the strange quark condensates, $\left\langle \bar{s}s\right\rangle$ do  not change significantly as a function of isospin asymmetric parameter of the  medium e.g., on shifting from symmetric to asymmetric nuclear medium, the magnitude of strange quark condensates, $\left\langle \bar{s}s\right\rangle$ change by 1$\%$ only,  at baryonic density, 4$\rho_B$, and in zero temperature situations. Also, in the same medium conditions, the percentage change of the mass and decay constant of $D_s (B_s)$ meson from its vacuum value is observed to be 3.3$\%$ (9.6$\%$) and 1.8$\%$ (30$\%$), respectively.  Likewise, in an isospin asymmetric matter, $I$=0.5, above values change to 3.1$\%$ (9.1$\%$) and 1.6$\%$ (29$\%$), respectively.

Among the various condensates present in the QCD sum rules (\cref{qcds}), the scalar strange quark condensates $\left\langle \bar{s}s\right\rangle$ has maximum contribution to the in-medium modification of $D_s$ and $B_s$ mesons. 
To understand this, in \cref{fig:individual} we plot the shift in mass of $D_s$  
(sublot(a)) and $B_s$ (subplot(b)) mesons as a function of squared Borel mass parameter, $M^2$. We observed that, if  all the condensates are set to zero, except $\left\langle \bar{s}s\right\rangle$, then the  mass of $D_s$($B_s$) meson was observed to be 1855(4472) MeV, and this can be compared with the mass 1840(4461) MeV, calculated in light of all the condensates, at baryonic density $4\rho_0$, in hot and symmetric hyperonic matter.  Also, we observe a negligible contribution of other condensates on the shift in mass of $D_s$ meson, if we set $\left\langle \bar{s}s\right\rangle$ = 0. In the present calculation, only the condensate $\left\langle s^\dag \iota D_o s \right\rangle$ is calculated in linear density approximation at higher density. However, it follows from \cref{fig:individual} that, the contribution of $\left\langle s^\dag \iota D_o s \right\rangle$ on the shift in masses and decay constants of $D_s/B_s$ mesons is negligible. Also, in this work, we evaluate the shift in masses and decay constants of $D_s/B_s$ mesons by taking the next to leading order term (NLO) to strange quark condensates $\left\langle \bar{s}s\right\rangle$. In \cref{table: indv}, we organize the numerical results of the shift in mass of $D_s$ meson, evaluated through next to leading order term (NLO), and  through the leading order term (LO) to $\left\langle \bar{s}s\right\rangle$. We observe the higher drop in the mass of $D_s$ meson, evaluated using the next to leading order term, as compared to its value calculated by taking the leading order term only.

Here, as mentioned earlier, while deriving the Borel transformed equation for pseudoscalar $D_s$ and $B_s$ mesons, we neglected the mass of strange quark $m_s$, i.e., ($m_c+m_s) \simeq m_c$.  However, if we consider the finite strange quark mass $m_s$, then we get some extra terms in the Borel transformed QCD sum rule equations $\propto$ $m_s\left\langle \bar{s}s\right\rangle$  as discussed in \cite{hilger2009, haya2004, lucha}.   
On following this analysis, we observe an increase in the masses and decay constants of the above mentioned mesons. However this increase is not much significant. For example, in symmetric nuclear medium, the  mass (decay constant) of $D_s$ meson increase by 0.2 $\%$ (0.25 $\%$) only at $\rho_B=\rho_0$ and $T$=0.

  Moreover, in the present work if we allow 10$\%$ change in the value of coupling constant $g_{{{D_s}N\Lambda_c}}$ $\approx$ $g_{{{D_s}N\Sigma_c}}$, then we observe a change in the magnitude of shift in mass of $D_s$ meson by 2$\%$ only. However, 10$\%$ change in the value of continuum threshold parameter $s_0$ cause significant change of 16$\%$, in the magnitude of shift in mass of $D_s$ meson. The other uncertainties of the results may be
the contribution of inelastic channels in scattering processes.

Furthermore, to check the reliability of our results at high density region, in \cref{fig:valid}, we plot the variation of strange quark condensates calculated through linear density approximation \cite{hilger2009}, as well as using chiral SU(3) model. In \cite{hilger2009}, using linear density approximation, authors calculated the strange quark condensates as,    $\left\langle \bar{s}s\right\rangle$ = 0.8$\left\langle \bar{q}q\right\rangle_{0}$+  $y\frac{\sigma_N \rho_B}{m_u + m_d}$, for $\sigma_N$ = 45 MeV and $m_u$+$m_d$ = 11 MeV. Here the term $\left\langle \bar{q}q\right\rangle_{0}$, is the vacuum value of light quark condensate and is given as (-0.245 GeV)$^3$. Also, the value of $y$ was taken to be 0.5. By considering these strange condensates only, in \cref{fig:valid} we plot the variation of the mass of $D_s(1968)$ meson as a function of baryonic density in symmetric nuclear medium.  Within linear density approximation, we observe linear decrease in $\left\langle \bar{s}s\right\rangle$, as well as in the mass of $D_s$ meson as a function of baryonic density. However, the decrease in $\left\langle \bar{s}s\right\rangle$, and of the in-medium mass is non-linear, if we evaluate $\left\langle \bar{s}s\right\rangle$  using the  chiral SU(3) model.  Similar response was observed in the case of light quark condensates $\left\langle \bar{q}q\right\rangle$ \cite{rahul2}. In \cite{rahul2} we noticed that, the non-linear behaviour of light quark condensates $\left\langle \bar{q}q\right\rangle$ calculated through chiral SU(3) model  is in accordance with the results of $\left\langle \bar{q}q\right\rangle$ calculated through loop contributions  beyond the linear density approximation \cite{kaiser}.

\begin{table}
\begin{tabular}{|l|l|l|l|l|l|l|l|l|l|l|}
\hline
& & \multicolumn{4}{c|}{$f_s$=0} & \multicolumn{4} {c|}{$f_s$=0.5}\\  \cline{3-10}

$D_s$& & \multicolumn{2}{c|}{T=0} &\multicolumn{2}{c|}{T=100MeV}& \multicolumn{2}{c|}{T=0} & \multicolumn{2}{c|}{T=100MeV}    \\\cline{3-10}

& & $\rho_0$ & $4\rho_0$  &$\rho_0$&4$\rho_0$&$\rho_0$&4$\rho_0$ & $\rho_0$ & 4$\rho_0$ \\ \hline
All Condensates  &NLO&-41&-64&-36 &-60 &-40 &-61 & -36 & -59\\ \cline{2-10}
  & LO &-30 &-43 & -22& -34 & -44& -80 & -38 & -70 \\ \cline{1-10}
$\left\langle s\bar{s} \right\rangle _N$ $\neq$ 0  &NLO &-37 &-53 & -28 &-42 &-56 &-100 &-49 &-88\\  \cline{2-10}
  &LO & -25 &-31 & -18&-23 &-39 &-65 &-33&-56\\  \hline
$\langle \bar{s} i D_0 s\rangle _N$ =0 &NLO&-40&-59&-30 &-47 &-60 & -108 & -52 &-95\\ \cline{2-10}
 & LO &-27 &-37 & -20 & -28 &-42  &-73  &-36 &-64\\ \hline

\end{tabular}
\caption{Contribution of individual condensates on the shift in mass of $D_s$ meson (in MeV) for two values of temperatures ($T$=0 and 100 MeV), strangeness fraction ($f_s$=0 and 0.5) and baryonic density ($\rho_0$ and 4$\rho_0$)  in symmetric medium ($I$=0).}
\label{table: indv}
\end{table}

\begin{figure}
\centering
\includegraphics[width=14cm,,height=12cm]{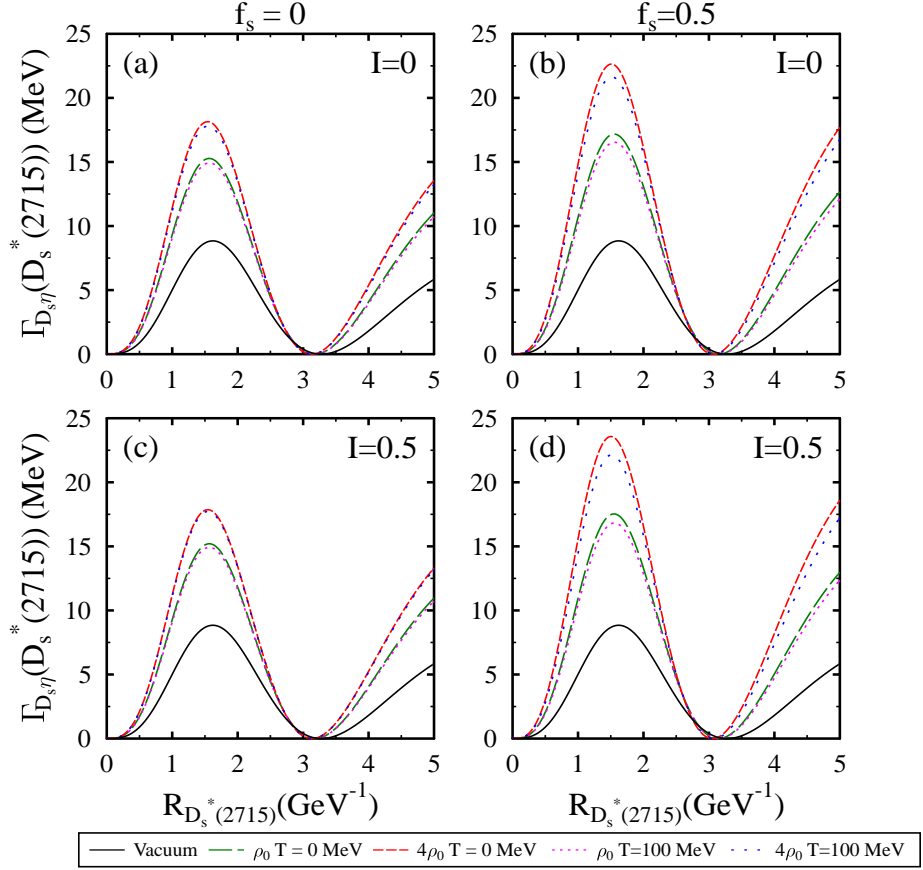}
\caption{Figure shows the variation of partial decay widths of $D_s^*(2715)$ mesons to a pairs of $D_s(1968)$ and $\eta$ mesons as a function of R$_{D_s^*(2715)}$(GeV$^-1$) for two values of baryonic densities ($\rho_0$ and $4\rho_0$), isospin asymmetric parameters ($I = 0$ and $0.5$), temperatures ($T$ = 0 and 100) MeV and strangeness fractions ($f_s = 0$ and $0.5$).} \label{fig:Ds(2715)}
\end{figure}

\begin{figure}
\centering
\includegraphics[width=14cm,,height=12cm]{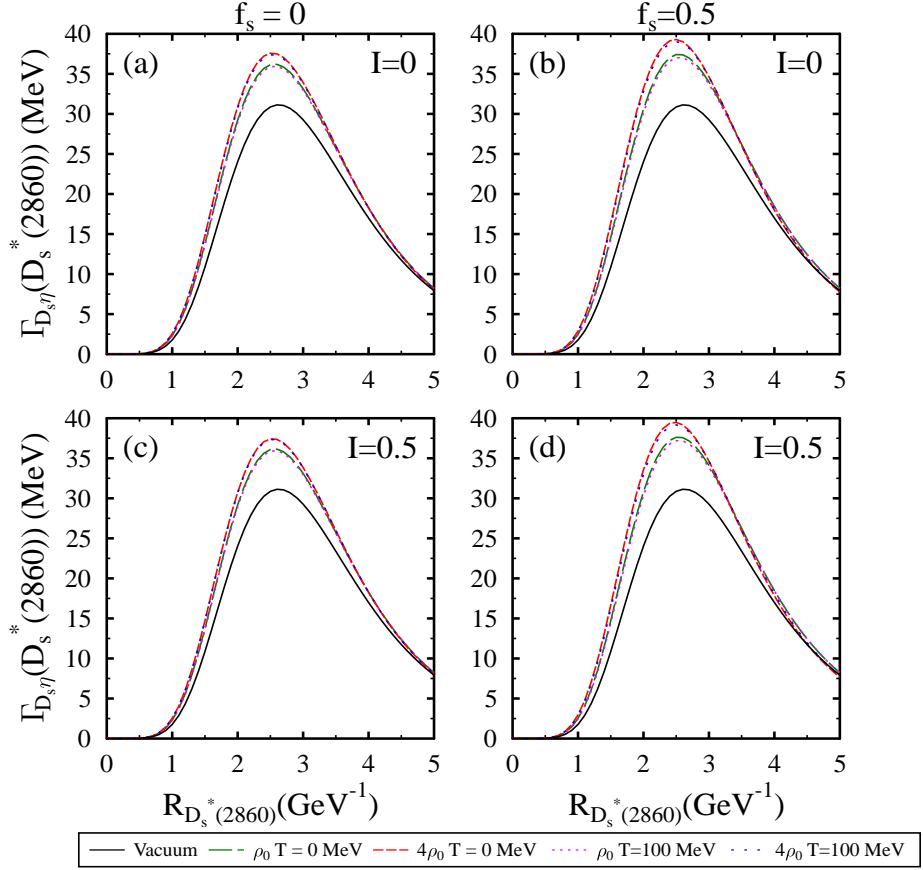}
\caption{Figure shows the variation of partial decay widths of $D_s^*(2860)$ mesons to a pairs of $D_s(1968)$ and $\eta$ mesons as a function of R$_{D_s^*(2860)}$ for two values of baryonic densities ($\rho_0$, $4\rho_0$), isospin asymmetric parameter ($I = 0$, $0.5$), temperatures ($T$ = 0, 100) MeV and strangeness fractions ($f_s = 0$, $0.5$)}\label{fig:Ds(2860)}
\end{figure}

\begin{figure}
\centering
\includegraphics[width=14cm,,height=12cm]{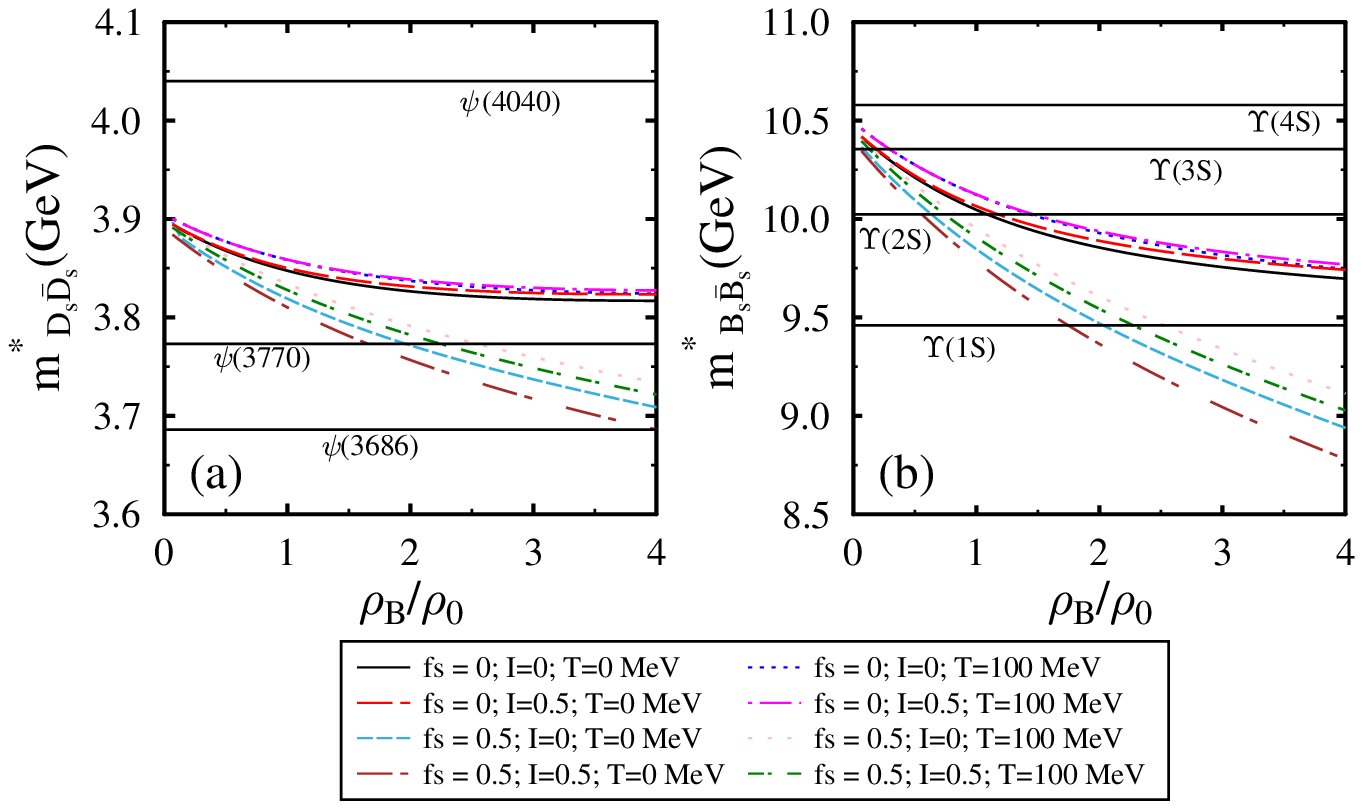}
\caption{We compare the masses of $D_s \bar{D_s}$ and  $B_s \bar{B_s}$ pairs, against charmonium (subplot(a)) and bottomonium (subplot(b)) states, respectively, for different medium situations, as described in the legend.} \label{fig:Ds_charm}
\end{figure}

 We now compare the results of the present investigation with the results of the other models. Using the self consistent coupled channel approach with t-dependent vector meson exchange deriving force \cite{CE,lutz1,hoff}, authors calculated the attractive interaction of $D_s$ meson in the hadronic matter. This is in accordance with the results of negative shift in the mass and hence favors the finite possibility of the formation of bound states of $D_s$ mesons with the nucleons as well as hyperons. 
 Quark meson coupling model, had been used in Ref. \cite{tsu}, and negative mass shift of about 60 MeV was observed for $B$ meson, at nuclear saturation density $\rho_0$.  
In Ref. \cite{yasup} using heavy quark
symmetry, the bound states of $DN$ and $BN$ states
were observed with the binding energies 1.4 MeV and 9.4 MeV, respectively. As in the present work we observe the attractive interaction for $D_s$ and $B_s$ mesons in the strange hadronic medium, and  similar attractive in-medium interactions were observed for $D$ and $B$ mesons in our previous work  \cite{rahul2} and therefore, one might expect the observation of   $D_sN$ and $B_sN$ bound states in future PANDA experiment of FAIR project. Recently, the chiral SU(3) model was generalized to SU(4) and was used to observe the effect of temperature, density and isospin asymmetric parameter on the in-medium masses of $D_s$ mesons \cite{pathak1}. Using the chiral SU(4) model authors observed the in-medium mass of $D_s$ meson as  1865(1875) MeV, at temperature $T$=0 ($T$=100 MeV), and at density, $\rho_B$=4$\rho_0$ in symmetric nuclear medium. Likewise, in asymmetric ($I$=0.5) nuclear medium the same values were observed to be 1875 and 1883 MeV for temperatures $T$=0 and $T$=100 MeV, respectively. However, in asymmetric hyperonic medium, at baryonic density 4$\rho_0$ because of non zero contribution of Weinberg-Tomozawa term, the mass degeneracy was broken and the values of masses of $D_s^+$ and $D_s^-$ were observed as 1859 and  1879 MeV respectively.

We can compare these values with our calculated values of 1842(1856)MeV at temperature $T$=0 ($T$=100 MeV) and baryonic density 4$\rho_0$, of  symmeitrc nuclear medium. In  \cite{pathak2}, $B_s$ mesons in asymmetric hyperonic medium were studied using the chiral effective approach generalized to the heavy quark sector. By using this method,  in asymmetric nuclear medium, at zero temperature, authors observed drop in the masses of $B_s^0$ and  $\bar{B_s^0}$ mesons of nearly 67(326) and 73(349) MeV, respectively at baryonic density $\rho_0$(4$\rho_0$). As was discussed earlier, in the present work, we calculate the average shift in mass of $B_s^0$($D_s^+$) and $\bar{B_s^0}$($D_s^-$) mesons by taking the averaged current density, as in \cref{tw}. We can compare these results with our calculated values of drop in the mass of $B_s$ meson, which are 285(515) MeV at baryon density $\rho_0$($4\rho_0$), zero temperature asymmetric nuclear matter.

In \cite{hilger2009}, the masses of $D/B$ and $D_s$ mesons were observed using the linear density approximate QCD sum rules, upto the normal nuclear density, $\rho_0$ only.  Using this analysis the authors observed the  mass splitting of $D_s-\bar{D_s}$ and  $D-\bar{D}$ as  well as of $B-\bar{B}$ mesons.  

In \cite{domi0708} authors discussed the temperature dependence of masses, decay constants and width of heavy pseudoscalar and vector mesons.
It was concluded in this work that the masses of pseudoscalar meson varies very slow with temperature, whereas decay constants decreases and become zero at temperature $T=T_c$.  We also note this slow increase in the masses and decay constants of $D_s$ and $B_s$ mesons as a function of temperature of the medium and this is due to fact that the scalar  fields $\sigma$ and $\zeta$ first increase very slowly as a function of temperature till a value of $T$ $\simeq$ 150 MeV, and beyond this value their is abrupt decrease\cite{ping2001}.

Now we will discuss the possible implication of negative shift in mass and decay constant of $D_s/B_s$ mesons in the hadronic matter. As discussed earlier, the results of the attractive interaction of $D_s$ and $B_s$ mesons in the hyperonic (along with nucleons) medium shows the finite possibility of the observation of bound states of $D_s$  and $B_s$ mesons with nucleons as well as with hyperons.  One may also expect that, this drop in the mass of $D_s$($B_s$) meson can enhance the decay channel of various excited charmonium(bottomonium) states and these higher sates may decay to $D_s\bar{D_s}(B_s\bar{B_s})$ states instead of $J/\psi$($\Upsilon$) states and may cause  $J/\psi$($\Upsilon$) suppression in the heavy ion collision experiments.
 For the better understanding, in \cref{fig:Ds_charm}, we compare the in-medium masses of $D_s\bar{D_s}$ pairs and $B_s\bar{B_s}$ pairs with the vacuum masses of various charmonium and bottomonium states, respectively. Here, we neglect the mass modification  of the charmonium and bottomonium states in the hadronic matter. It follows from \cref{fig:Ds_charm} that, the drop in the masses of $D_s$ and $B_s$ mesons can open up the decay channel  of the type $A \to D_s\bar{D_s}$ and $A \to B_s\bar{B_s}$, where $A$ can be considered as some excited charmonium or bottomonium state. As these higher charmonium and bottomonium states are considered as a major source of $J/\psi$ and $\Upsilon$ states, therefore the drop in the mass of $D_s$ and $B_s$ mesons may cause these higher states to decay to the pairs of $D_s\bar{D_s}$ and $B_s\bar{B_s}$ states instead of $J/\psi$ and $\Upsilon$ states, respectively and hence this may suppress the production of $J/\psi$ and $\Upsilon$ states in heavy ion collision experiments.  Moreover, one may also expect a decrease in the production yield of the higher charmonium and bottomonium states, and  the new decay channel may cause change in the total decay width of these higher states and this might also help in the future HIC experiments to measure the total decay width of various hidden charm and bottom states.  
 
Further, the $D_s$ and $B_s$ mesons have large leptonic and semileptonic decay widths \cite{dilepton}. As the masses of these mesons drop in the medium, one can also expect enhanced production of $D_s$ and $B_s$ mesons. Therefore, as mentioned in ref. \cite{pathak2}, this may  also cause an increase in the dilepton spectra observed in the heavy ion collision experiments. However, if we consider the leptonic decay width of a particular $D_s$ meson, given as \cite{pgupta,heis}
\begin{align}
\Gamma_{(D_s \to l \nu)} = \frac{G_F^2}{8\pi} f_{D_s}^{*2} |V_{cq}|^2 m_l^2 \left(1-\frac{m_l^2}{m_{D_s}^{*2}}\right)^2 m_{D_s}^*,
\label{eq_leptonic1}
\end{align}
and use medium modified values of masses and decay constants of $D_s$ mesons in the above equation, then we observe a decrease in the value of leptonic 
decay width.
In \cref{eq_leptonic1}, $G_F$ is the Fermi coupling constant =1.1663787 $\times$ $10^{-5}$ GeV$^{-2}$, $m_l$ is the lepton mass, $V_{cq}$ is the CKM matrix  = 0.97 \cite{kimbook}, $m_{D_s}^*$ and $f_{D_s}^*$ are the in-medium mass and decay constant of $D_s$ meson, respectively.
For example, at baryonic density, $\rho_B$ = 4$\rho_0$, in symmetric strange hadronic ($f_s$=0.5) matter, we observe the leptonic decay width of a particular channel ($D_s$ $\to$ $\mu \bar{\nu_\mu}$) as 5.3419 $\times$ $10^{-9}$ keV, which is small as compared to its vacuum value 6.3355 $\times$ $10^{-9}$ keV. 
 Similarly, for the leptonic decay, $D_s$  $\to$ $e \bar{\nu_e}$, for the same medium situations,  the value of leptonic width will be 1.2735 $\times$ $10^{-13}$ keV and this is also small as compared to its vacuum value 1.5091 $\times$ $10^{-13}$ keV.  
  Therefore, we argue that the medium modification of leptonic decay widths can't be neglected and one may also expect that the drop in the mass of $D_s$ meson may decrease the leptonic yield in the HIC experiments. This may contradicts our previously mentioned point of enhanced dilepton spectra in the HIC experiments which is due to enhanced production of $D_s$ mesons. However we argue that to understand the exact dilepton spectra, more work is required in the field of medium modification of leptonic decay widths and also on the production yield of $D_s$ mesons in HIC experiments. To make a definite conclusion on the temperature and density dependence of these mesons, as well as to validate the theoretical models, we seek for the possible outcomes of the future experiments like CBM and PANDA, at GSI Germany.
 \subsection{In-medium partial decay width of $D_s^*$(2715) and $D_s^*$(2860) mesons}
 \label{sub_decay} 
 In this section, using $^3P_0$ model, we shall calculate the in-medium partial decay width of 
$D^*_s(2715)$ and $D_s^*(2860)$ for the processes
$D_s^*(2715)$  $\to$ $D_s(1968)$ + $\eta$ and $D_s^*(2860)$  $\rightarrow$ $D_s(1968)$ + $\eta$, respectively (\cref{fig:Ds(2715)} and \cref{fig:Ds(2860)}). 
The open charm states $D^*_s(2715)$ and $D_s^*(2860)$, are reported by Belle and Babar Collaborations, respectively. We consider $D_s^*(2715)$ and $D_s^*(2860)$ states  as $2 ^3S_1$  and $1 ^3D_1$ candidate \cite{liu,liu1,godfrey}.
 As mentioned earlier, for calculating the partial decay width we shall consider the medium modified masses of daughter products only. We shall use the in-medium mass of $D_s(1968)$ meson calculated in \cref{sub_mass}. 
Additionally, we include the medium modified mass of $\eta$ meson, calculated using heavy-baryon chiral perturbation theory combined with relativistic meanfield theory \cite{zhong1}. In \cite{zhong1} authors studied the in-medium mass of $\eta$ mesons  in symmetric nuclear matter at zero temperature  including next to leading order term. Following this work, we used $-75$ and $-120$ MeV as values of mass shift for $\eta$ mesons at $\rho_0$ and $4\rho_0$, respectively.
As no work is still available on the study of mass shift
of $\eta$ mesons in asymmetric strange matter at finite temperatures,
we used same above values in asymmetric strange matter.

    In the present work, we notice a significant influence of the  modification in the mass of $D_s$ and $\eta$ mesons on the partial decay width, in addition to the respective $R_A$ values. Furthermore, we observe  the enhanced in-medium values of decay widths of both the processes  $D_s^*(2715)$  $\rightarrow$ $D_s(1968)$ + $\eta$ and $D_s^*(2860)$  $\rightarrow$ $D_s(1968)$ + $\eta$, as compared to their vacuum values.  For example, from \cref{fig:Ds(2715)} and \cref{fig:Ds(2860)} we notice that for any constant value of temperature, whether in symmetric ($I$=0) or in asymmetric ($I$=0.5), nuclear medium ($f_s$=0), the values of the decay widths at the nuclear matter denser medium $\rho_0$, are more than their vacuum values. Also, we observe that on increasing the density of the medium ($\rho_0$ $\rightarrow$ 4$\rho_0$), decay widths further increase. This can be understood on the basis that, the finite baryonic density causes decrease in the mass of the $D_s(1968)$ and $\eta$ mesons, which further enhances the decay channel. We also observe nodes in $\Gamma_{D_s \eta} (D_s^*(2715))$ at a particular value of $R_A$ = 3.27 GeV$^{-1}$. To understand the occurrence of node we recall that on solving  the spatial integral analytically for the decay process ($D_s^*(2715)$  $\rightarrow$ $D_s(1968)$ + $\eta$), we get \cite{liu}  
\begin{align} I^{0,0}_{0,0}&=& - \sqrt{\frac{1}{2}}
\frac{|\mathbf{k}_{B}|
    \pi^{1/4}R_{A}^{3/2}R_{B}^{3/2}R_{C}^{3/2}}
{(R_{A}^{2}+R_{B}^{2}+R_{C}^{2})^{5/2}}             \Bigg\{
-6(R_{A}^{2}+R_{B}^{2}+R_{C}^{2})(1+\xi)+R_A^2  \bigg[
4+20\xi\nonumber\\&&+\mathbf{k}_{B}^2
(R_{A}^{2}+R_{B}^{2}+R_{C}^{2}) (-1+\xi)^2(1+\xi) \big]
 \Bigg\}
\exp\bigg[-\frac{\mathbf{k}_{B}^{2}R_{A}^{2}(R_{B}^{2}+R_{C}^{2})}{8(R_{A}^{2}+R_{B}^{2}+R_{C}^{2})}\bigg].
\label{express-3}
\end{align}

 In above, the polynomial part written within braces become zero as the value of $R_{D_s^*(2715)}$ approaches 3.27 GeV$^{-1}$ and thus,  nodes are observed in $\Gamma_{D_s \eta} (D_s^*(2715))$ at this value of $R_A$. This happens because of the nodal structure of the simple harmonic oscillator wave function of the first excited state $1^-(2^3 S_1)$ of the parent meson.    Furthermore, as the masses of $D_s(1968)$/$\eta$ mesons contribute through the momentum ${\bf K} = {\bf k_B}$, given by \cref{K} and it is present in the polynomial part of \cref{express-3}, and therefore, the modification in the masses of $D_s(1968)$ and $\eta$ mesons have finite (though small) effect  on the position of node. It is clear from \cref{K} that the decrease in the masses of $D_s(1968)$/$\eta$ mesons cause increase in the value of momentum ${\bf K}$. This increase in the value of momentum ${\bf K}$, further change the position of node e.g., at zero temperature, for baryonic density $\rho_B$ = 4$\rho_0$, $\eta = 0.5$ and $f_s = 0.5$, the position of node shift to $R_A$ = 3.07 GeV$^{-1}$, compared to the original value $R_A$=3.27 GeV$^{-1}$.  
  On the other hand,  we do not observe  nodes in the decay process, $ D_s^*(2860) \to {D_s(1968) + \eta}$, and this is because of the state $1^3 D_1$ of the parent meson. Further, for any constant value of baryonic density, isospin asymmetric parameter and temperature of the medium, we observe an enhancement in the decay width as we move from nuclear ($f_s$=0) to hyperonic medium ($f_s$=0.5). 
As explained in  \cref{sub_mass}, the mass of $D_s$ meson decrease further in the strange medium ($f_s$=0.5), and this further cause increase in the possibility of $D_s^*(2715)$ and $D_s^*(2860)$ mesons decaying to channel ($D_s(1968)$ $\eta$).  On the other hand, as mentioned in \cref{sub_mass}, finite temperature causes an increase in the mass of $D_s(1968)$ meson and this further causes decrease in the decay width of above mentioned processes.   
Talking about the isospin asymmetric parameter, as discussed earlier, the strange quark condensate is not much sensitive to the isospin asymmetry  of the medium. Therefore,  we could not observe much significant effect of isospin asymmetry on both the values $\Gamma_{D_s \eta} (D_s^*(2715))$ and $\Gamma_{D_s \eta} (D_s^*(2860))$, in nuclear medium. However, in cold ($T$=0) hyperonic medium ($f_s$=0.5),  and four times the nuclear matter density (4$\rho_0$), we observe a slight increase in the decay width of the above mentioned decay processes, as is clear from  \cref{fig:Ds(2715)} and \cref{fig:Ds(2860)}. 
This is due to  decrease in the mass of $D_s(1968)$ meson in asymmetric hyperonic matter (mentioned in \cref{sub_mass}).  
Now to compare the results, as far as our knowledge about the literature is concerned,  in-medium decay width of $D_s^*(2715)$ and $D_s^*(2860)$ mesons have not been studied, however many authors have predicted their spectroscopy and calculated their decay widths in vacuum  using microscopic $^3P_0$ model \cite{godfrey,liu,close,close1,Li1,Li}.
Apart from $^3P_0$ model other models like quark model including quark meson effective Lagrangian approach and heavy quark effective theories had also been applied to investigate the decay widths of excited $D_s$ mesons \cite{zhong,chen}.
Here in the present work, we take a non relativistic transition operator to calculate the partial decay width of above mentioned mesons\cite{micu,yao,barn1,barn2,close,sego,ferre1,ferre2,ferre3,close1,zhong,chen,Li1}.   However, if we consider a relativistic approach to formulate $^3P_0$  model as mentioned in  \cite{akleh},  then there will be  interaction Hamiltonian containing Dirac quark fields, 
$H_I$ = g $\int_{a}^{b} d^3x \bar{\psi} \psi$, for g = 2$\gamma m_s$. We fix the value of coupling constant $g$ through the original value of quark pair creation strength parameter $\gamma = 6.4$. Now considering relativistic correction to the mass of strange quark $m_s$, if we increase its value by 5$\%$, then we observe a decrease in the partial decay width of  $\Gamma_{D_s \eta} (D_s^*(2715))$($\Gamma_{D_s \eta} (D_s^*(2860))$) by 10$\%$(5$\%$), to its original values in symmetric matter at baryon density $\rho_0$ and temperature $T$ = 0. 
 
\begin{figure}
\centering
\includegraphics[width=17cm,,height=15cm]{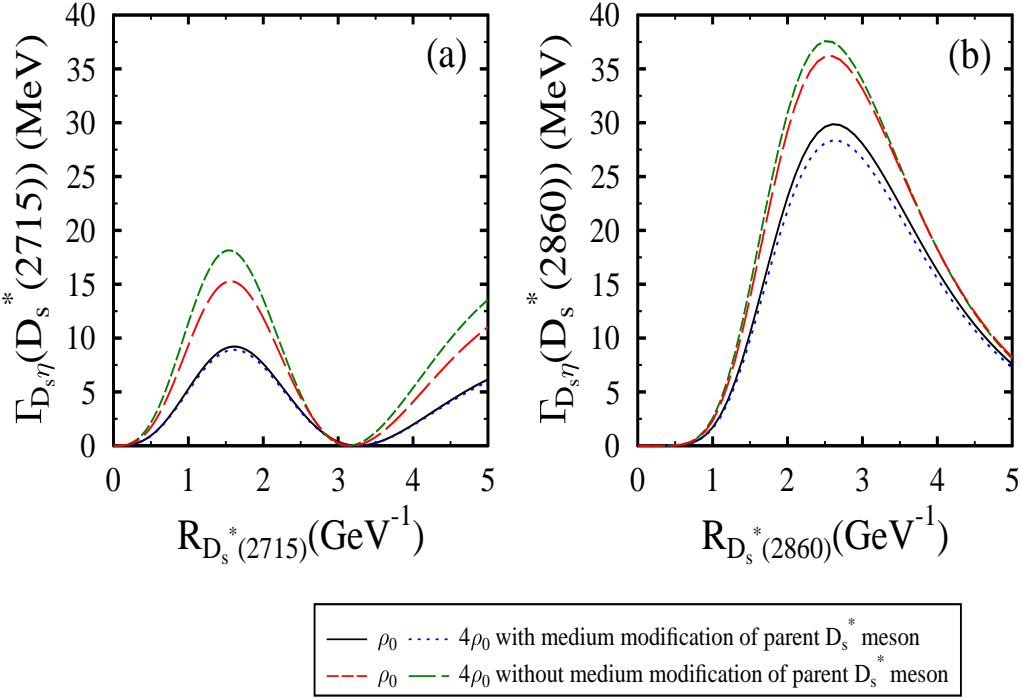}
\caption{Variation of partial decay width of the  $D_s^*(2715)$ and $D_s^*(2860)$ mesons as a function of respective $R_A$ values with and without including the medium modified masses of  $D_s^*(2715)$ and $D_s^*(2860)$ mesons mesons.}\label{fig8:Dsmass}
\end{figure}

In the above discussion of in-medium partial decay width, we neglected the mass modification of parent   $D_s^*(2715)$ and $D_s^*(2860)$ mesons. As far as our knowledge is concern, no work is available for   in-medium study of masses of these parent mesons. However, in order to understand the effect of medium modified masses of $D_s^*(2715)$ and $D_s^*(2860)$ mesons on their partial decay width, we recall the in-medium study of the ground state strange vector mesons $D_s^*$ and $B_s^*$ \cite{Rahul}.  In ref. \cite{Rahul}, in symmetric nuclear medium, at $\rho_B=\rho_0$ and $T = 0$, we  observed a decrease in the values of masses of $D_s^*$ and $B_s^*$ mesons by 2.3$\%$ and 3.7$\%$, respectively, from their vacuum values. 
Likewise at higher baryonic density 4$\rho_0$, medium mass of  the above mentioned mesons further decrease to  3.7$\%$ and 6.8$\%$, respectively from their vacuum values.
 From this experience of the medium modification of the ground state vector $c\bar{s}$ and $\bar{b}s$ states, we also expect a similar kind of drop  in the masses of excited $c\bar{s}$ states.  
 Here it should be noted that, as a function of baryonic density, the in-medium mass of $D^*_s$ meson decreases very fast till baryonic density  $\rho_B \simeq 1.5\rho_0$, and beyond that this decrease become slow.  This behavior can be understood on the basis of the in-medium behavior of light quark and gluon condensates, which further depend on the in-medium behavior of $\sigma$, $\zeta$ and $\chi$ fields (discussed in previous section). The decrease in the values of $\sigma$ and $\zeta$ fields as a function of baryonic density is very fast till the baryonic density $\rho_B$ $\simeq$ 1.5$\rho_0$ and beyond that the decrease become slow. From this we expect a decrease in the mass of $D_s^*(2715)$ and $D_s^*(2860)$ mesons at baryonic density $\rho_0$(4$\rho_0$) of approximately 3$\%$(5$\%$).
In \cref{fig8:Dsmass}, we present the effect of medium modified masses of $D_s^*(2715)$ and $D_s^*(2860)$ mesons on their partial decay widths. We observe that the inclusion of the  medium modified masses of these parent mesons further decrease  the value of the partial decay width. For example, if we allow 3$\%$ decrease in the masses of $D_s^*(2715)$ and $D_s^*(2860)$ mesons, in symmetric nuclear medium, at $\rho_B = \rho_0$ and $T = 0$, then the values of  $\Gamma_{D_s \eta} (D_s^*(2715))$ and $\Gamma_{D_s \eta} (D_s^*(2860))$ are observed to be 8 and 17.2 MeV, respectively. Likewise, expecting 5$\%$ drop in the mass of parent mesons $\rho_B = 4\rho_0$, we observed partial decay widths as 7.6 and 16.2 MeV, respectively. Clearly, these values are less than the respective values of $\Gamma_{D_s \eta} (D_s^*(2715))$ and $\Gamma_{D_s \eta}(D_s^*(2860))$   observed without modifying the mass of parent mesons, i.e.,  12 and 30 MeV (13 and 32 MeV), in symmetric matter at baryon density $\rho_0$ (4$\rho_0$) and  temperature $T = 0$.
 Expecting more  large shift in the mass of excited $D_s^*$ mesons,
 say, 6$\%$ at $\rho_0$, and 10$\%$ at $4\rho_0$,  we observed further decrease in the values of the partial decay width.

More detailed study of masses of  $D_s^*(2715)$ and $D_s^*(2860)$ mesons in asymmetric strange hadronic medium and its possible implication on the in-medium decay width of above discussed processes will be a goal of future study.

 \section{Summary}
 \label{summary} We observed the negative shift in masses and decay constants of the pseudoscalar $D_s(1968)$ and $B_s(5370)$ mesons, using the chiral SU(3) model and the QCD sum rules technique.  Furthermore, we take the in-medium mass of $D_s(1968)$ meson as an application in $^3P_0$ model and observe the in-medium partial decay widths of excited $c\bar{s}$ states, i.e., $D_s^*(2715)$ and $D_s^*(2860)$ mesons decaying to $D_s(1968)+\eta$. We observe that, as the mass of $D_s(1968)$ meson decrease in the hyperonic (along with the nucleons) medium, this results in the significant increase in the corresponding partial decay widths. 
In the present paper, we have neglected the modification in masses of parent mesons $D_s^*(2715)$ and $D_s^*(2860)$.  
We emphasis on the fact that, even at higher density region,
the above calculated in-medium decay width never seems to be more than the vacuum decay width of the  $D_s^*(2715)$/$D_s^*(2860)$ to the channel $DK$.
In our future it will be of interest to include the other decay channel in the calculations to evaluate the in-medium decay width.

\acknowledgements
 The authors gratefully acknowledge the financial support from
the Department of Science and Technology (DST), Government of India for research project under
 fast track scheme for young scientists (SR/FTP/PS-209/2012).

\end{document}